\documentclass[11pt,a4paper]{article}

\usepackage{amssymb}
\usepackage{graphicx}
\usepackage{bm}

\usepackage{mathrsfs}
\usepackage{cite}

\begin{document}

\title{Finiteness of the triple gauge-ghost vertices in ${\cal N}=1$ supersymmetric gauge theories: the two-loop verification}

\author{M.D.Kuzmichev\,${}^{ab}$, N.P.Meshcheriakov\,${}^{ab}$, S.V.Novgorodtsev\,${}^a$,\\ V.V.Shatalova\,${}^{ac}$, I.E.Shirokov\,${}^a$,
K.V.Stepanyantz\,${}^{ab}$ $\vphantom{\Big(}$
\medskip\\
{\small{\em Moscow State University, Faculty of Physics,}}\\
${}^a${\small{\em Department of Theoretical Physics,}}\\
${}^b${\small{\em Department of Quantum Theory and High Energy Physics,}}\\
{\small{\em 119991, Moscow, Russia}}\\
\vphantom{1}\vspace*{-2mm}\\
${}^c${\small{\em Moscow State University, AESC MSU – Kolmogorov boarding school}},\\
{\small{\em Department of Physics}}\\
{\small{\em 119192, Moscow, Russia}}\\
\vphantom{1}\vspace*{-2mm}}

\maketitle

\begin{abstract}
By an explicit calculation we demonstrate that the triple gauge-ghost vertices in a general renormalizable ${\cal N}=1$ supersymmetric gauge theory are UV finite in the two-loop approximation. For this purpose we calculate the two-loop divergent contribution to the $\bar c^+ V c$-vertex proportional to $(C_2)^2$ and use the finiteness of the two-loop contribution proportional to $C_2 T(R)$ which has been checked earlier. The theory under consideration is regularized by higher covariant derivatives and quantized in a manifestly ${\cal N}=1$ supersymmetric way with the help of ${\cal N}=1$ superspace. The two-loop finiteness of the vertices with one external line of the quantum gauge superfield and two external lines of the Faddeev--Popov ghosts has been verified for a general $\xi$-gauge. This result agrees with the nonrenormalization theorem proved earlier in all orders, which is an important step for the all-loop derivation of the exact NSVZ $\beta$-function.
\end{abstract}

\unitlength=1cm

\section{Introduction}
\hspace*{\parindent}

In supersymmetric quantum field theory models ultraviolet (UV) divergences are essentially restricted by some nonrenormalization theorems. Cancellations of these divergences in theories with extended supersymmetry are especially impressive. They lead to the all-loop finiteness of ${\cal N}=4$ supersymmetric Yang--Mills theory \cite{Sohnius:1981sn,Grisaru:1982zh,Howe:1983sr,Mandelstam:1982cb,Brink:1982pd} and to the finiteness of ${\cal N}=2$ supersymmetric gauge theories beyond the one-loop approximation \cite{Grisaru:1982zh, Howe:1983sr,Buchbinder:1997ib}. If a gauge group and a representation for the matter superfields are chosen in such a way that the one-loop $\beta$-function vanishes \cite{Banks:1981nn}, then it is possible to construct ${\cal N}=2$ models finite in all loops \cite{Howe:1983wj}. The finiteness persists even after adding some special terms breaking extended supersymmetry \cite{Parkes:1982tg,Parkes:1983ib,Parkes:1983nv}.

Although in the phenomenologically interesting ${\cal N}=1$ supersymmetric theories divergences appear in all loops, the nonrenormalization theorems exist even in this case. Namely, the superpotential does not receive divergent quantum corrections \cite{Grisaru:1979wc}, the $\beta$-function is related to the anomalous dimension of the matter superfields by the NSVZ equation \cite{Novikov:1983uc,Jones:1983ip,Novikov:1985rd,Shifman:1986zi}, and the triple gauge-ghost vertices are finite in all loops \cite{Stepanyantz:2016gtk}.
Wonderfully, it is even possible to construct all-loop finite ${\cal N}=1$ supersymmetric theories \cite{Parkes:1984dh,Kazakov:1986bs,Ermushev:1986cu,Lucchesi:1987he,Lucchesi:1987ef}. Certainly, the nonrenormalization theorems and, in particular, the NSVZ equation are very important for this purpose, see, e.g., \cite{Heinemeyer:2019vbc,Stepanyantz:2021dus}. Also the NSVZ equation is very useful for investigating the fixed points in ${\cal N}=1$ supersymmetric theories \cite{Seiberg:1994pq,Ryttov:2017khg}. The NSVZ-like equations describing the renormalization of the gaugino mass \cite{Hisano:1997ua,Jack:1997pa,Avdeev:1997vx} can also be considered as nonrenormalization theorems for the terms which softly break supersymmetry. Using these and other relations between the renormalization of soft breaking parameters and the renormalization of the rigid theory \cite{Hisano:1997ua,Jack:1997pa,Avdeev:1997vx,Kazakov:1998bz,Kazakov:1998uj,Jack:1998iy,Jack:1997eh,Kazakov:1997nf} it is possible to construct all-loop finite theories with softly broken supersymmetry \cite{Jack:1997eh,Kazakov:1997nf,Kazakov:1995cy}.

However, it should be noted that most nonrenormalization theorems hold only for certain renormalization prescriptions. Even the finiteness of the superpotential is in general valid only for the superspace quantization, while in the case of using the component formulation one should specially tune a subtraction scheme. In fact \cite{Buchbinder:2014wra}, the finiteness of ${\cal N}=2$ supersymmetric theories takes place in the case of using the manifestly ${\cal N}=2$ supersymmetric quantization with the help of ${\cal N}=2$ harmonic superspace \cite{Galperin:1984av,Galperin:2001uw,Buchbinder:2001wy} and an invariant regularization \cite{Buchbinder:2015eva}. In the component formulation three-loop divergences in ${\cal N}=2$ supersymmetric theories have been revealed by explicit calculations \cite{Avdeev:1982np,Velizhanin:2008rw}.

Also it was found that (starting from the order which corresponds to the three-loop $\beta$-function and the two-loop anomalous dimension of the matter superfields) the NSVZ equation in ${\cal N}=1$ supersymmetric gauge theories does not hold in the $\overline{\mbox{DR}}$-scheme \cite{Jack:1996vg,Jack:1996cn,Jack:1998uj,Harlander:2006xq,Mihaila:2013wma}, when a theory is regularized by dimensional reduction \cite{Siegel:1979wq} and divergences are removed by modified minimal subtraction \cite{Bardeen:1978yd}. Nevertheless, it is possible to make a specially tuned finite renormalization of the gauge coupling constant which restores the NSVZ equation \cite{Jack:1996vg,Jack:1996cn,Jack:1998uj,Harlander:2006xq,Mihaila:2013wma}, and this possibility is nontrivial, because the NSVZ equation leads to some scheme independent consequences \cite{Kataev:2013csa,Kataev:2014gxa}.

An all-loop prescription giving a scheme in which the Abelian NSVZ equation \cite{Vainshtein:1986ja,Shifman:1985fi} is valid has been constructed with the help of the higher covariant derivative regularization \cite{Slavnov:1971aw,Slavnov:1972sq,Slavnov:1977zf} in the supersymmetric version \cite{Krivoshchekov:1978xg,West:1985jx}. This regularization allows revealing some interesting features of the quantum correction structure \cite{Stepanyantz:2019lyo}, for example, the factorization of integrals giving the $\beta$-function into integrals of (double) total derivatives \cite{Soloshenko:2003nc,Smilga:2004zr,Pimenov:2009hv,Stepanyantz:2011bz,Stepanyantz:2011jy,Stepanyantz:2019ihw}. In the case of using this regularization the NSVZ equation turns out to be valid for the renormalization group functions (RGFs) defined in terms of the bare coupling constant \cite{Stepanyantz:2011jy,Stepanyantz:2014ima} in all loops for an arbitrary renormalization prescription.\footnote{For theories regularized by dimensional reduction this is not true, see Refs.~\cite{Aleshin:2015qqc,Aleshin:2016rrr} for details.} This statement has been confirmed by explicit calculations (see, e.g., \cite{Stepanyantz:2011jy,Stepanyantz:2012zz,Kazantsev:2014yna,Aleshin:2020gec}) in the three-loop approximation, where the scheme dependence becomes essential, and was used for constructing simple all-loop prescriptions giving NSVZ schemes for the standard RGFs (defined in terms of the renormalized coupling constant). In particular, in all loops the HD+MSL \cite{Kataev:2013eta} and on-shell \cite{Kataev:2019olb} schemes belong to the continuous set of the NSVZ schemes \cite{Goriachuk:2018cac} in the Abelian case. (By definition, in the HD+MSL scheme a theory is regularized by Higher Derivatives supplemented by Minimal Subtraction of Logarithms \cite{Shakhmanov:2017wji,Stepanyantz:2017sqg}, when for removing divergences only powers of $\ln\Lambda/\mu$ are included into renormalization constants.\footnote{Minimal subtractions of logarithms can be used with various versions of the higher derivative regularization, so that there is continuous set of the HD+MSL schemes, all of them being NSVZ.}) For the softly broken supersymmetric electrodynamics the exact NSVZ-like equation describing the renormalization of the photino mass is also valid in the HD+MSL scheme \cite{Nartsev:2016nym,Nartsev:2016mvn}.

Generalization of the results obtained for ${\cal N}=1$ supersymmetric quantum electrodynamics (SQED) to the non-Abelian case is not trivial.\footnote{The only exception is the NSVZ-like equation for the Adler $D$-function \cite{Adler:1974gd} in ${\cal N}=1$ SQCD which can be derived by the same method as the Abelian NSVZ equation, see \cite{Shifman:2014cya,Shifman:2015doa} and also \cite{Kataev:2017qvk} for details.} Surprisingly, it turned out that a key statement required for this is the nonrenormalization theorem for the triple gauge-ghost vertices in which the external lines correspond to the quantum gauge superfield, the Faddeev--Popov ghost, and the Faddeev--Popov antighost. For the superfield formulation of ${\cal N}=1$ supersymmetric gauge theories in an arbitrary $\xi$-gauge it has been derived in Ref.~\cite{Stepanyantz:2016gtk} with the help of the Slavnov--Taylor identities \cite{Taylor:1971ff,Slavnov:1972fg} and rules for calculating supergraphs \cite{Gates:1983nr,West:1990tg,Buchbinder:1998qv}. The all-loop finiteness of the triple gauge-ghost vertices allowed to rewrite the NSVZ equation in an equivalent form \cite{Stepanyantz:2016gtk} (see Ref.~\cite{Korneev:2021zdz} for a generalization to the case of theories with multiple gauge couplings). It relates the $\beta$-function to the anomalous dimensions of quantum superfields (namely, of the quantum gauge superfield, of the Faddeev--Popov ghosts, and of the matter superfields). This new form of the NSVZ equation was subsequently derived by perturbative methods in all loops for RGFs defined in terms of the bare couplings for theories regularized by higher covariant derivatives \cite{Stepanyantz:2019ihw,Stepanyantz:2020uke} (see also \cite{Stepanyantz:2019lfm}). Consequently, for RGFs defined in terms of the renormalized couplings, the HD+MSL prescription gives some NSVZ schemes. (As in the Abelian case, all HD+MSL schemes constitute a continuous set, see  Refs. \cite{Goriachuk_Conference,Goriachuk:2020wyn,Korneev:2021zdz}.) The validity of both original and new forms of the NSVZ equation has been confirmed by numerous multiloop calculations \cite{Stepanyantz:2019lyo,Pimenov:2009hv,Stepanyantz:2011cpt,Shakhmanov:2017soc,Kazantsev:2018nbl,Kuzmichev:2019ywn}. Knowing in which scheme the NSVZ equation is valid, one can obtain a $\beta$-function by calculating the anomalous dimension of the matter superfields in the previous loops, see, e.g., the derivation of the three-loop $\beta$-function for a general renormalizable ${\cal N}=1$ supersymmetric theory in Ref.~\cite{Kazantsev:2020kfl}.

Thus, the nonrenormalization of the triple gauge-ghost vertices appears to be an important statement needed for understanding structure of quantum corrections in supersymmetric theories. That is why it is very desirable to verify it by explicit perturbative calculations. One of similar statements known earlier for some theories formulated in terms of usual fields in the Landau gauge $\xi\to 0$ \cite{Dudal:2002pq,Capri:2014jqa} has been verified by an explicit four-loop calculation in Ref.~\cite{Chetyrkin:2004mf}. For ${\cal N}=1$ supersymmetric theories and a general $\xi$-gauge the one-loop finiteness of the triple gauge-ghost vertices has explicitly been checked in \cite{Stepanyantz:2016gtk}. The two-loop verification was started in Ref.~\cite{Kuzmichev:2021yjo}, where it has been demonstrated that the two-loop contribution to the considered vertices proportional to $C_2 T(R)$ is UV finite. However, except for this contribution (which comes from the two-loop superdiagrams containing a loop of the matter superfields) there is a contribution proportional to $(C_2)^2$, which will be considered in this paper. Below we will demonstrate that it is also finite in the UV region. This will complete the two-loop verification of the nonrenormalization theorem for the triple gauge-ghost vertices.

The paper is organized as follows. In Sect.~\ref{Section_2} we recall various aspects of the superfield formulation, (higher derivative) regularization, and quantization of ${\cal N}=1$ supersymmetric gauge theories. The nonrenormalization theorem for the triple gauge-ghost vertices is formulated in Sect.~\ref{Section_3}. Also in this section we discuss how one can simplify loop calculations that are required to verify it. Such a calculation made in the two-loop approximation is described in Sect.~\ref{Section_4}, where we explicitly demonstrate the UV finiteness of the triple gauge-ghost vertices.

\section{The theory under consideration: action, regularization, and quantization}
\hspace*{\parindent}\label{Section_2}

It is convenient to investigate $\mathcal{N}=1$ supersymmetric gauge theories using the superspace formulation because in this case $\mathcal{N}=1$ supersymmetry is manifest. At the classical level the action of the theory with a simple gauge group $G$ and chiral matter superfields $\phi_i$ in a representation $R$ is given by the expression

\begin{eqnarray}\label{Action_Classical_Superfield}
	&& S = \frac{1}{2 e_0^2}\,\mbox{Re}\,\mbox{tr}\int d^4x\,
	d^2\theta\,W^a W_a + \frac{1}{4} \int d^4x\, d^4\theta\,\phi^{*i}
	(e^{2V})_i{}^j \phi_j\nonumber\\
	&&\qquad\qquad\qquad\qquad\qquad\qquad  +
	\bigg\{\int d^4x\,d^2\theta\,\Big(\frac{1}{4} m_0^{ij} \phi_i \phi_j + \frac{1}{6}\lambda_0^{ijk} \phi_i
	\phi_j \phi_k\Big) + \mbox{c.c.}\bigg\},\qquad
\end{eqnarray}

\noindent
which is written in terms of $\mathcal{N}=1$ superfields. Here $e_0$ and $\lambda_0^{ijk}$ stand for the bare gauge and Yukawa couplings, respectively, and $m_0^{ij}$ is a (symmetric) bare mass matrix. The strength of the Hermitian gauge superfield $V$ (again at the classical level) is defined as $W_a \equiv \bar D^2(e^{-2V} D_a e^{2V})/8$. In the action (\ref{Action_Classical_Superfield}) the gauge superfield is expanded as $V = e_0 V^A t^A$ inside $W_a$ and as $V=e_0 V^A T^A$ in the second term. In our conventions $t^A$ and $T^A$ denote the generators of the fundamental representation and of the representation $R$, respectively, and satisfy the equations

\begin{equation}
	\mbox{tr}(t^A t^B) = \frac{1}{2}\delta^{AB};
	\qquad
	\mbox{tr}(T^A T^B) = T(R) \delta^{AB}.
\end{equation}

\noindent
The generators of the adjoint representation are written as $(T_{Adj}^A)_{B}{}^{C} = -i f^{ABC}$, where $f^{ABC}$ are the structure constants. The similar equation for them defines the quadratic Casimir $C_2$,

\begin{equation}
	\mbox{tr}(T^A_{Adj} T^B_{Adj}) = f^{ACD} f^{BCD} \equiv C_2 \delta^{AB}.
\end{equation}

If the masses and Yukawa couplings satisfy the conditions

\begin{eqnarray}\label{Condition_Masses_for_gauge_inv}
	&& m_0^{ik} (T^A)_k{}^{j} + m_0^{kj} (T^A)_k{}^{i} = 0;\vphantom{\Big(}\\
	\label{Condition_Yukawa_for_gauge_inv}
	&& \lambda_0^{ijm} (T^A)_m{}^{k} + \lambda_0^{imk} (T^A)_m{}^{j} +
	\lambda_0^{mjk} (T^A)_m{}^{i} = 0,\vphantom{\Big(}
\end{eqnarray}

\noindent
the action (\ref{Action_Classical_Superfield}) is invariant under the gauge transformations

\begin{equation}\label{Gauge_Transformations}
	\phi \to e^{A}\phi;\qquad e^{2V} \to e^{-A^+} e^{2V} e^{-A},
\end{equation}

\noindent
parameterized by a chiral gauge superfield $A$ which takes values in the Lie algebra of the gauge group. If we also require that

\begin{equation}\label{Mass_Squared}
	m_0^{ik} m^*_{0kj} = m_0^2 \delta_j^i,
\end{equation}

\noindent
\noindent
then (see \cite{Kuzmichev:2021yjo} for details) the generators will also satisfy the condition

\begin{equation}\label{Anomaly_Cancellation_Condition}
	\mbox{tr}\Big(T^A \{T^B, T^C\}\Big) = 0,
\end{equation}

\noindent
which ensures the cancellation of the gauge anomalies \cite{Bertlmann:1996xk,Ohshima:1999jg}. Below we will always assume that this condition is valid.

Quantizing gauge theories it is convenient to use the background field method \cite{DeWitt:1965jb,Abbott:1980hw,Abbott:1981ke}, which allows constructing the effective action manifestly invariant under the (background) gauge transformations. In the supersymmetric case \cite{Gates:1983nr,Grisaru:1982zh} the background-quantum splitting is nonlinear. Moreover, one should take into account that the quantum gauge superfield is renormalized in a nonlinear way \cite{Piguet:1981fb,Piguet:1981hh,Tyutin:1983rg}, due to which this splitting is introduced with the help of the replacement

\begin{equation}\label{Splitting}
	e^{2V} \to e^{2{\cal F}(V)} e^{2\bm{V}}.
\end{equation}

\noindent
In this equation $\bm{V}$ stands for the Hermitian background gauge superfield, and $V$ hereafter denotes the quantum gauge superfield (which obeys the constraint $V^+ = e^{-2\bm{V}} V e^{2\bm{V}}$). The function ${\cal F}(V) = e_0 {\cal F}(V)^A t^A$ (or ${\cal F}(V) = e_0 {\cal F}(V)^A T^A$) includes various nonlinear terms which are really needed for making the renormalization in the supersymmetric case. In particular, in the lowest approximations it can be written as \cite{Juer:1982fb,Juer:1982mp}

\begin{equation}\label{Function_F}
	{\cal F}(V)^A = V^A + e_0^2\, y_0\, G^{ABCD} V^B V^C V^D + \ldots,
\end{equation}

\noindent
where  $ G^{ABCD} \equiv \big(f^{AKL} f^{BLM} f^{CMN} f^{DNK} + \mbox{permutations of $B$, $C$, and $D$}\big)/6$ is a totally symmetric tensor and $y_0$ is a new bare parameter. The calculation of the two-loop anomalous dimension of the Faddeev--Popov ghosts made in Ref.~\cite{Kazantsev:2018kjx} has demonstrated that without taking into account the renormalization of $y_0$ the renormalization group equations cannot be satisfied. Therefore, in this paper the nonlinear terms in the function ${\cal F}(V)$ are also very essential. (Note that according to \cite{Piguet:1984mv} this function can contain only odd powers of $V$.)

After the replacement (\ref{Splitting}) we introduce the higher covariant derivative regularization by adding a term with higher covariant derivatives to the action. The resulting regularized action can be presented in the form

\begin{eqnarray}\label{Action_Regularized}
	&& S_{\mbox{\scriptsize reg}} = \frac{1}{2 e_0^2}\,\mbox{Re}\, \mbox{tr} \int d^4x\, d^2\theta\, W^a \Big[e^{-2\bm{V}} e^{-2{\cal F}(V)}\,  R\Big(-\frac{\bar\nabla^2 \nabla^2}{16\Lambda^2}\Big)\, e^{2{\cal F}(V)}e^{2\bm{V}}\Big]_{Adj} W_a \qquad\nonumber\\
	&& + \frac{1}{4} \int d^4x\,d^4\theta\, \phi^{*i} \Big[\, F\Big(-\frac{\bar\nabla^2 \nabla^2}{16\Lambda^2}\Big) e^{2{\cal F}(V)}e^{2\bm{V}}\Big]_i{}^j \phi_j
	+ \bigg\{ \int d^4x\,d^2\theta\, \Big(\frac{1}{4} m_0^{ij} \phi_i \phi_j \qquad\nonumber\\
	&& + \frac{1}{6} \lambda_0^{ijk} \phi_i \phi_j \phi_k\Big) + \mbox{c.c.} \bigg\},\qquad
\end{eqnarray}

\noindent
where

\begin{equation}
	W_a = \frac{1}{8}\bar D^2 \left(e^{-2\bm{V}} e^{-2{\cal F}(V)} D_a (e^{2{\cal F}(V)} e^{2\bm{V}})\right),
\end{equation}

\noindent
and higher powers of the covariant derivatives

\begin{equation}\label{Covariant_Derivative_Definition}
	\nabla_a = D_a;\qquad \bar\nabla_{\dot a} = e^{2{\cal F}(V)} e^{2\bm{V}} \bar D_{\dot a} e^{-2\bm{V}} e^{-2{\cal F}(V)}
\end{equation}

\noindent
are present in the regulator functions $R(x)$ and $F(x)$. Both these functions should rapidly grow at infinity and should be equal to $1$ at $x=0$. The parameter $\Lambda$ has the dimension of mass and effectively plays a role of an UV cutoff. Also in Eq.~(\ref{Action_Regularized}) we use the notation

\begin{equation}
	P(x)_{Adj} Q = (P_0 + P_1 x + P_2 x^2 + \ldots)_{Adj} Q \equiv P_0 Q + P_1 [x,Q] + P_2 [x,[x,Q]] +\ldots
\end{equation}

It is convenient to fix a gauge without breaking the background gauge invariance. This can be done by adding the gauge fixing term

\begin{equation}\label{Action_Gauge_Fixing}
	S_{\mbox{\scriptsize gf}} = -\frac{1}{16\xi_0 e_0^2}\, \mbox{tr} \int d^4x\, d^4\theta\,  \bm{\nabla}^2 V K\Big(-\frac{\bm{\bar\nabla}^2 \bm{\nabla}^2}{16\Lambda^2}\Big)_{Adj} \bm{\bar\nabla}^2 V
\end{equation}

\noindent
(which depends on a real bare parameter $\xi_0$) to the regularized action. This term includes one more regulator function $K(x)$ (with the same asymptotics at $x=0$ and $x \to \infty$ as the functions $R(x)$ and $F(x)$) and the background covariant derivatives

\begin{equation}
	\bm{\nabla}_a \equiv D_a; \qquad \bm{\bar\nabla}_{\dot a} \equiv e^{2\bm{V}} \bar D_{\dot a} e^{-2\bm{V}}.
\end{equation}

\noindent
The gauge fixing procedure also prescribes adding the ghost actions

\begin{eqnarray}
	\label{Action_Faddeev-Popov_Ghosts}
	&& S_{\mbox{\scriptsize FP}} = \frac{1}{2} \int
	d^4x\,d^4\theta\, \frac{\partial {\cal F}^{-1}(\widetilde V)^A}{\partial {\widetilde V}^B}\left.\vphantom{\frac{1}{2}}\right|_{\widetilde V = {\cal F}(V)} \left(e^{2\bm{V}}\bar c e^{-2\bm{V}} +
	\bar c^+ \right)^A\nonumber\\
	&&\qquad\qquad\qquad\quad \times \left\{\vphantom{\frac{1}{2}} \smash{\Big(\frac{{\cal F}(V)}{1-e^{2{\cal F}(V)}}\Big)_{Adj} c^+
		+ \Big(\frac{{\cal F}(V)}{1-e^{-2{\cal F}(V)}}\Big)_{Adj}
		\Big(e^{2\bm{V}} c e^{-2\bm{V}}\Big)}\right\}^B;\qquad\\
	&& \vphantom{1}\nonumber\\
	\label{Action_Nielsen-Kallosh_Ghosts}
	&&  S_{\mbox{\scriptsize NK}} = \frac{1}{2e_0^2}\,\mbox{tr} \int d^4x\,d^4\theta\, b^+ \Big(K\Big(-\frac{\bm{\bar\nabla}^2 \bm{\nabla}^2}{16\Lambda^2}\Big) e^{2\bm{V}}\Big)_{Adj} b,
\end{eqnarray}

\noindent
where $c$ and $\bar c$ stand for the Faddeev--Popov ghost and antighost superfields, respectively, and $b$ denotes the Nielsen--Kallosh ghost superfield. All these superfields are chiral and anticommuting.

To regularize the one-loop divergences which survive even after introducing the higher derivative terms, we insert the Pauli--Villars determinants into the generating functional. According to Refs.~\cite{Aleshin:2016yvj,Kazantsev:2017fdc}, in the supersymmetric case one can use two such determinants. The first one, needed for regularizing the one-loop (sub)divergences produced by the gauge and ghost superfields, has the form

\begin{equation}
	\mbox{Det}(\mbox{PV},M_\varphi)^{-1} = \int D\varphi_1 D\varphi_2 D\varphi_3 \exp(i S_\varphi),
\end{equation}

\noindent
where the functional integration is performed over three commuting chiral superfields $\varphi_1$, $\varphi_2$, and $\varphi_3$ belonging to the adjoint representation of the gauge group. Their action is written as

\begin{eqnarray}
	&& S_\varphi = \frac{1}{2e_0^2} \mbox{tr}\int d^4x\, d^4\theta\, \Big(\varphi_1 ^+ \Big[ R\Big(-\frac{\bar\nabla^2 \nabla^2}{16\Lambda^2}\Big)e^{2{\cal F}(V)}  e^{2\bm{V}}\Big]_{Adj}\varphi_1 + \varphi_2^+ \Big[ e^{2{\cal F}(V)} e^{2\bm{V}}\Big]_{Adj}\varphi_2\qquad\nonumber\\
	&& + \varphi_3^+ \Big[ e^{2{\cal F}(V)} e^{2\bm{V}}\Big]_{Adj}\varphi_3\Big) + \frac{1}{2e_0^2}\Big(\mbox{tr}\int d^4x\, d^2\theta\, M_\varphi (\varphi_1^2 + \varphi_2^2 + \varphi_3^2) +\mbox{c.c.}\Big).\qquad
\end{eqnarray}

\noindent
The second Pauli--Villars determinant removes the (sub)divergences coming from a matter loop. It can be presented as the functional integral over the commuting chiral superfields $\Phi_i$ in a certain representation $R_{\mbox{\scriptsize PV}}$,

\begin{equation}
	\mbox{Det}(\mbox{PV},M)^{-1} = \int D\Phi \exp(iS_\Phi),
\end{equation}

\noindent
where

\begin{equation}
	S_\Phi = \frac{1}{4} \int d^4x\, d^4\theta\, \Phi^+ F\Big(-\frac{\bar\nabla^2 \nabla^2}{16\Lambda^2}\Big)e^{2{\cal F}(V)} e^{2\bm{V}} \Phi
	+ \Big(\frac{1}{4}\int d^4x\, d^2\theta\, M^{ij} \Phi_i \Phi_j +\mbox{c.c.}\Big).
\end{equation}

\noindent
Note that the representation $R_{\mbox{\scriptsize PV}}$ should admit the gauge invariant mass term such that $M^{ik} M^*_{kj} = M^2 \delta_j^i$. Consequently \cite{Kuzmichev:2021yjo}, its generators satisfy the condition

\begin{equation}\label{Anomaly_Cancellation_Condition_PV}
	\mbox{tr}\Big(T^A_{\mbox{\scriptsize PV}} \{T^B_{\mbox{\scriptsize PV}}, T^C_{\mbox{\scriptsize PV}}\}\Big) = 0
\end{equation}

\noindent
analogous to Eq.~(\ref{Anomaly_Cancellation_Condition}).

We will always assume that the masses of the Pauli--Villars superfields are proportional to the dimensionful cutoff parameter $\Lambda$ of the higher derivative regularization,

\begin{equation}
	M_\varphi \equiv a_\varphi \Lambda; \qquad M \equiv a \Lambda,
\end{equation}

\noindent
the coefficients $a_\varphi$ and $a$ being independent of couplings.

After inserting the Pauli--Villars determinants the generating functional of the regularized theory takes the form

\begin{eqnarray}
	&& Z[\bm{V},\mbox{Sources}] = \int D\mu\, \Big(\mbox{Det}(\mbox{PV},M)\Big)^{T(R)/T(R_{\mbox{\tiny PV}})} \,\,\, \mbox{Det}(\mbox{PV},M_{\varphi})^{-1} \qquad\nonumber\\
	&&\qquad\qquad\qquad\qquad\qquad\qquad\qquad \times \exp\Big(i S_{\mbox{\scriptsize reg}} + i S_{\mbox{\scriptsize gf}} + i S_{\mbox{\scriptsize FP}}
    + i S_{\mbox{\scriptsize NK}} + i S_{\mbox{\scriptsize sources}}\Big),\qquad
\end{eqnarray}

\noindent
where the integration measure is denoted by $D\mu$ and  all relevant sources are included into the term $S_{\mbox{\scriptsize sources}}$. The effective action $\Gamma$ is constructed with the help of the standard prescription, as a Legendre transform of the generating functional for the connected Green functions $W\equiv -i\ln Z$.

\section{The nonrenormalization theorem for the triple gauge-ghost vertices}
\label{Section_3}
\hspace*{\parindent}

In this paper we investigate the triple gauge-ghost vertices. In such vertices three external lines correspond to the Faddeev--Popov antighost, the quantum gauge superfield, and the Faddeev--Popov ghost. There are four types of these vertices corresponding to $\bar c V c$, $\bar c^+ V c$, $\bar c V c^+$, and $\bar c^+ V c^+$ external lines. The tree contributions to them are encoded in the expression

\begin{equation}\label{Vertex:3-gauge-ghost}
\frac{ie_0}{4} f^{ABC} \int d^4x\,d^4\theta\, (\bar c^A + \bar c^{+A}) V^B (c^C + c^{+C}),
\end{equation}

\noindent
which is easily derived from the Faddeev--Popov action (\ref{Action_Faddeev-Popov_Ghosts}). At the quantum level with the help of dimensional and chirality considerations we can present the $\bar c^+ V c$ and $\bar c^+ V c^+$ vertices in the form

\begin{eqnarray}\label{Gamma_Three_Point_Contribution1}
	&&\hspace*{-5mm} \Delta\Gamma_{\bar c^+ V c} = \frac{i e_0}{4} f^{ABC} \int d^4\theta\, \frac{d^4p}{(2\pi)^4} \frac{d^4q}{(2\pi)^4} \bar c^{+A}(p+q,\theta)\Big(s(p,q) \partial^2\Pi_{1/2}V^B(-p,\theta)\nonumber\\
	&&\hspace*{-5mm} \qquad\qquad\qquad\quad\ \ \, + {\cal S}_\mu(p,q) (\gamma^\mu)_{\dot a}{}^{b} D_b \bar D^{\dot a} V^B(-p,\theta) + {\cal S}(p,q) V^B(-p,\theta)\Big) c^C(-q,\theta);\qquad\\
	\label{Gamma_Three_Point_Contribution2}
	&&\hspace*{-5mm} \Delta\Gamma_{\bar c^+ V c^+} = \frac{i e_0}{4} f^{ABC} \int d^4\theta\, \frac{d^4p}{(2\pi)^4} \frac{d^4q}{(2\pi)^4} \bar c^{+A}(p+q,\theta) \widetilde {\cal S}(p,q) V^B(-p,\theta) c^{+C}(-q,\theta),\qquad
\end{eqnarray}

\noindent
where $\partial^2\Pi_{1/2} \equiv - D^a \bar D^2 D_a/8$ is the supersymmetric transversal projection operator. The structure of two remaining vertices is similar to that above.

According to the nonrenormalization theorem proved in \cite{Stepanyantz:2016gtk} (see also \cite{Korneev:2021zdz} for its extension to the case of theories with multiple gauge couplings) the triple gauge-ghost vertices are UV finite in all orders. This statement should be valid for theories formulated in ${\cal N}=1$ superspace and for a general $\xi$-gauge, unlike some earlier analogs \cite{Dudal:2002pq,Capri:2014jqa} valid for theories formulated in terms of the usual fields in the Landau gauge $\xi\to 0$.

The nonrenormalization of the triple gauge-ghost vertices imposes some restrictions to the renormalization constants of the theory. In our notation the renormalization constants $Z_\alpha$, $Z_c$, and  $Z_V$ for the gauge coupling constant $\alpha = e^2/4\pi$, for the Faddeev--Popov ghosts, and for the quantum gauge superfield, respectively, are defined as

\begin{equation}\label{Renormalization_Constants}
	\frac{1}{\alpha_0} = \frac{Z_\alpha}{\alpha};\qquad \bar c^A c^B = Z_c\, \bar c_R^A c_R^B; \qquad V^A = Z_V V_R^A,
\end{equation}

\noindent
where the subscript $R$ stands for renormalized superfields. All triple gauge-ghost vertices have the same renormalization constant $Z_\alpha^{-1/2} Z_c Z_V$. Therefore, their finiteness leads to the equation

\begin{equation}
	\frac{d}{d\ln \Lambda} \left( Z_\alpha^{-1/2} Z_c Z_V \right) =0,
\end{equation}

\noindent
so that it is possible to choose a subtraction scheme (e.g., HD+MSL) in which

\begin{equation}
Z_\alpha^{-1/2} Z_c Z_V  =1.
\end{equation}

The one-loop contributions to the functions  $s$, ${\cal S}_\mu$, ${\cal S}$, and $\widetilde {\cal S}$ have been calculated in Ref.~\cite{Stepanyantz:2016gtk}. The resulting expressions are UV finite, although individual superdiagrams are logarithmically divergent. The two-loop verification of the nonrenormalization theorem was initiated in Ref.~\cite{Kuzmichev:2021yjo}, where it was proved that the contribution proportional to $C_2T(R)$ is finite in the UV region. Certainly, to check this it is sufficient to investigate only one of the considered vertices, for example, the $ \bar{c}^+ V c$ vertex given by Eq.~(\ref{Gamma_Three_Point_Contribution1}) because the renormalization constants for all of them coincide. Note that the functions $s$ and ${\cal S}_\mu$ are determined by integrals with a negative superficial degree of divergence (namely, $-2$ and $-1$ in units of mass, respectively). Therefore, the divergent contributions to these functions can only arise from divergent subgraphs of the corresponding superdiagrams. However, for the considered renormalizable theories these subdivergences are eliminated by the renormalization in the previous orders.

In this paper using the same reasoning we will verify the finiteness of the remaining two-loop contribution to the triple gauge-ghost vertices which is proportional to $(C_2)^2$. Again we consider the $\bar{c}^+ V c$ vertex and  calculate the corresponding two-loop contribution to the function ${\cal S}$. It comes from the superdiagrams in which all internal lines correspond to superfields in the adjoint representation, i.e., to the quantum gauge superfield, the Faddeev--Popov ghosts, and the Pauli--Villars superfields $\varphi_{1,2,3}$. Following Ref.~\cite{Kuzmichev:2021yjo}, to extract the expression for the function ${\cal S}$, in superdiagrams contributing to the $\bar c^+ V c$ vertex we substitute the quantum gauge superfield on the external leg by the expression $\bar D^2 H$, where $H$ is a Hermitian superfield,

\begin{equation}\label{Substitution_V}
	V\to \bar D^2 H.
\end{equation}

\noindent
Certainly, this replacement is formal, because the right hand side is not Hermitian. Nevertheless, after it a part of the effective action corresponding to the vertex (\ref{Gamma_Three_Point_Contribution1}) takes the form

\begin{equation}\label{Gamma_cD2Hc}
    \Delta\Gamma_{\bar c^+ V c}\Big|_{V\to \bar D^2 H} = \frac{i e_0}{4} f^{ABC} \int d^4\theta\, \frac{d^4p}{(2\pi)^4} \frac{d^4q}{(2\pi)^4} \bar c^{+A}(p+q,\theta) {\cal S}(p,q) \bar D^2 H^B(-p,\theta) c^C(-q,\theta)
\end{equation}

\noindent
and contains only the function ${\cal S}$. To simplify the calculations, we study the superdiagrams contributing to this expression in the limit of the vanishing external momenta $p \to 0$, $q \to 0$, exactly as in Ref.~\cite{Kuzmichev:2021yjo}. This is possible because the terms proportional to the external momenta have a negative superficial degree of divergence and the corresponding subdivergences are removed by the renormalization in the previous orders. Therefore, the vanishing result in the limit $p \to 0$, $q \to 0$ ensures the finiteness of the considered contribution in the UV region.

\section{Two-loop contributions to the triple gauge-ghost vertices}
\label{Section_4}

\subsection{Two-loop contribution proportional to $C_2 T(R)$}
\hspace*{\parindent}

As we have already discussed, supergraphs which contribute to the triple gauge-ghost vertices in the two-loop approximation can naturally be divided into two parts. The first one contains superdiagrams with a loop of the matter superfields or a loop of the Pauli--Villars superfields $\Phi_i$ and is proportional to the factor $C_2 T(R)$. The second contribution consists of the superdiagrams in which all internal lines correspond to the quantum gauge superfield, the Faddeev--Popov ghosts, and the Pauli--Villars superfields $\varphi_{1,2,3}$. This contribution is proportional to $(C_2)^2$.

The sum of all superdiagrams contributing to the $C_2 T(R)$ part of the $\bar c^+ V c$ vertex has been calculated in Ref. \cite{Kuzmichev:2021yjo}, where it was demonstrated that it is finite in the UV region (for theories which satisfy the anomaly cancellation condition (\ref{Anomaly_Cancellation_Condition})). Therefore, to verify the finiteness of the triple gauge-ghost vertices, we need to investigate only the contribution proportional to $(C_2)^2$. This will be done in the next subsection.

\subsection{Two-loop contribution proportional to $(C_2)^2$}
\hspace*{\parindent}

To analyze the cancellation of UV divergences, it is convenient to split the two-loop superdiagrams contributing to the $(C_2)^2$ part of the considered vertex into five groups.

\begin{figure}[h]
	\begin{picture}(0,6.7)
		\put(1,5.4){\includegraphics[scale=0.14]{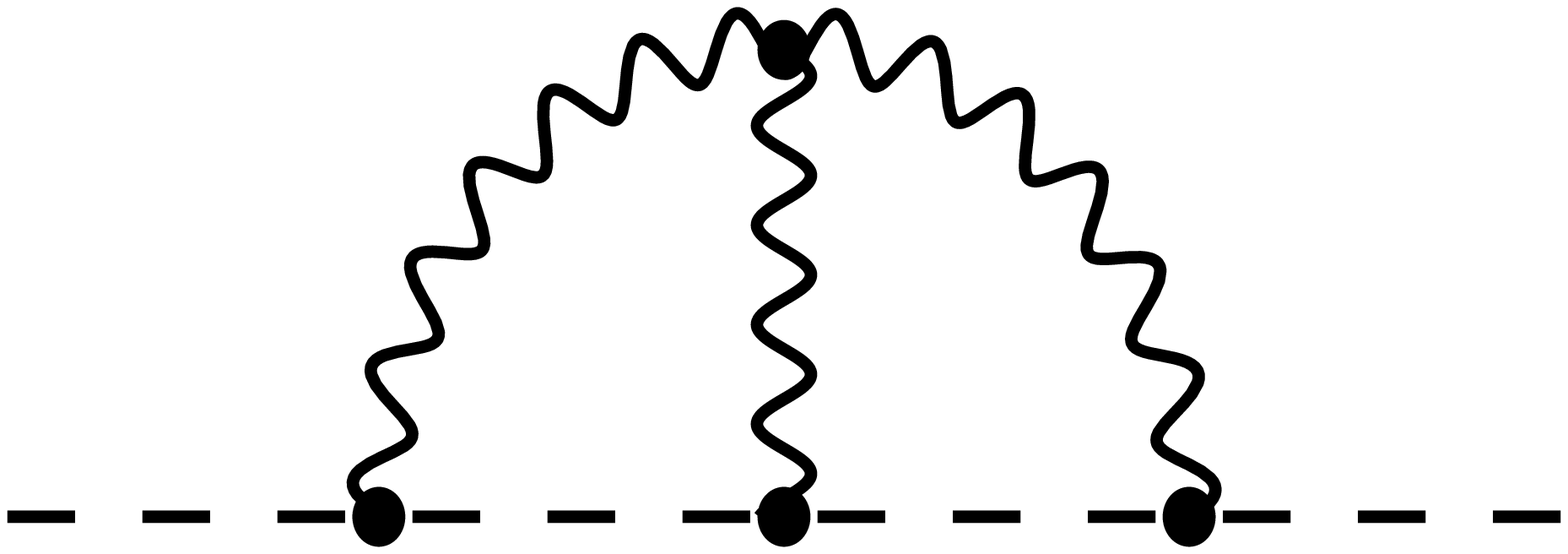}}
		\put(1,6.5){(a1)}
		\put(1,5){$ \bar{c}^+ $}
		\put(3.4,5){$ c$}
		
		\put(5,5.4){\includegraphics[scale=0.14]{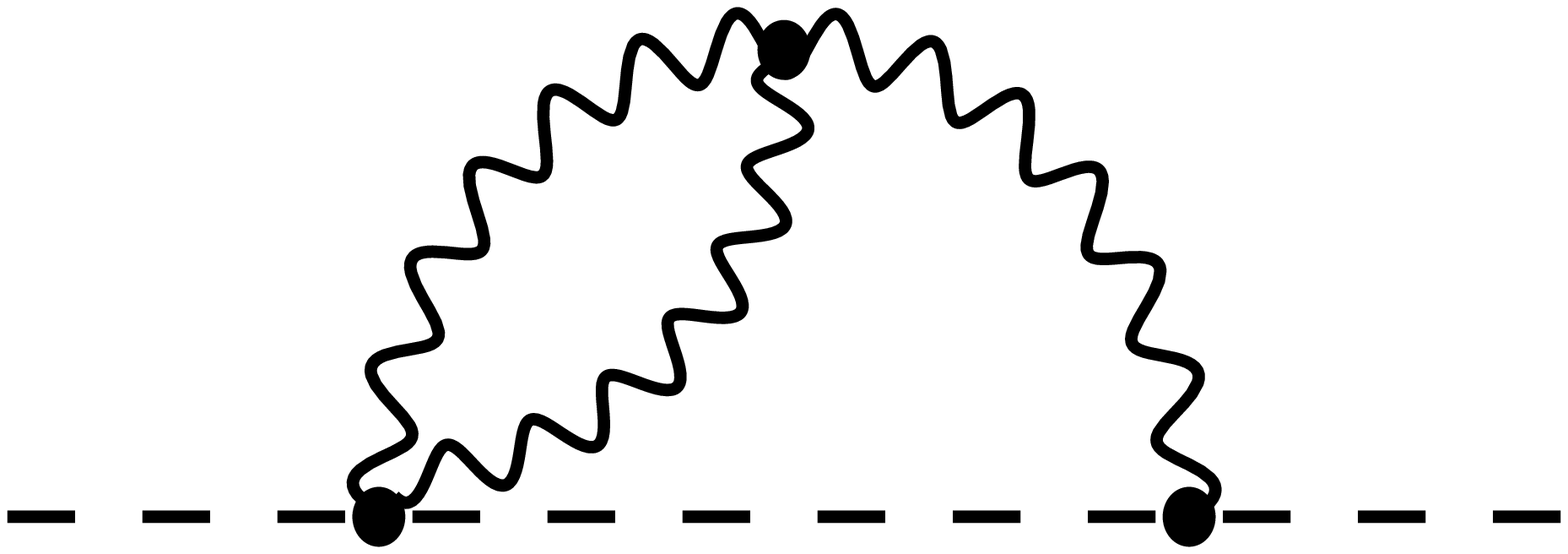}}
		\put(5,6.5){(a2)}
		\put(9,5.4){\includegraphics[scale=0.14]{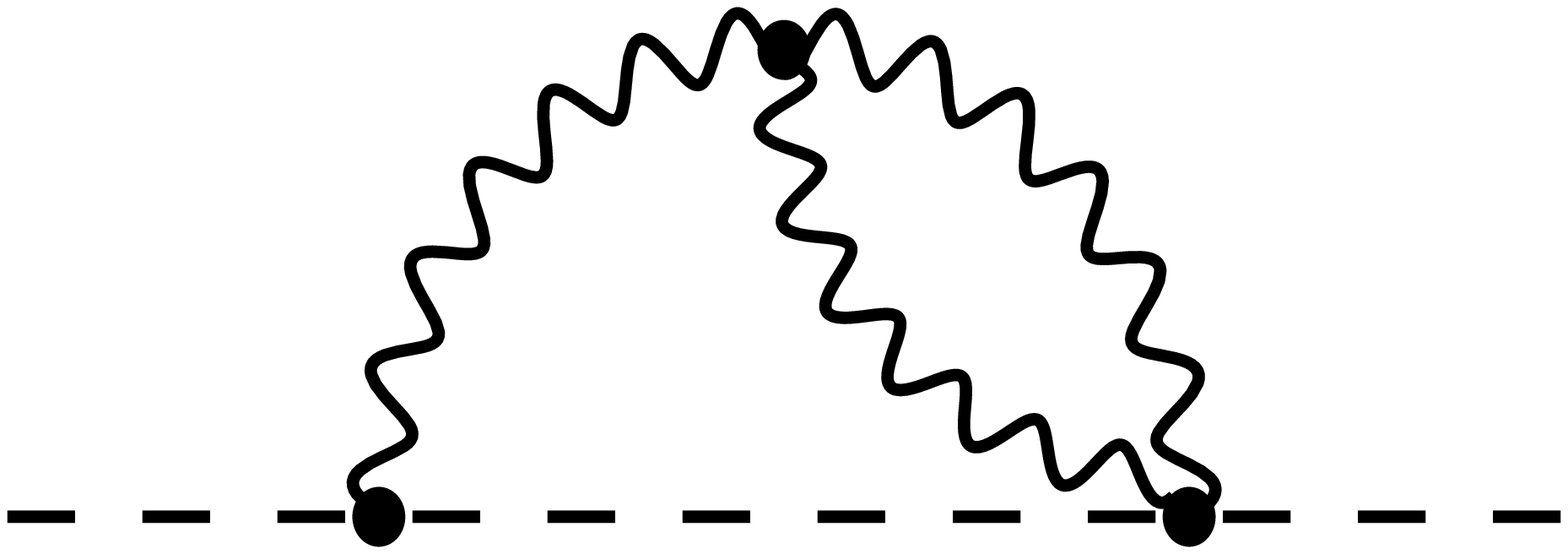}}
		\put(9,6.5){(a3)}
		\put(13,4.95){\includegraphics[scale=0.14]{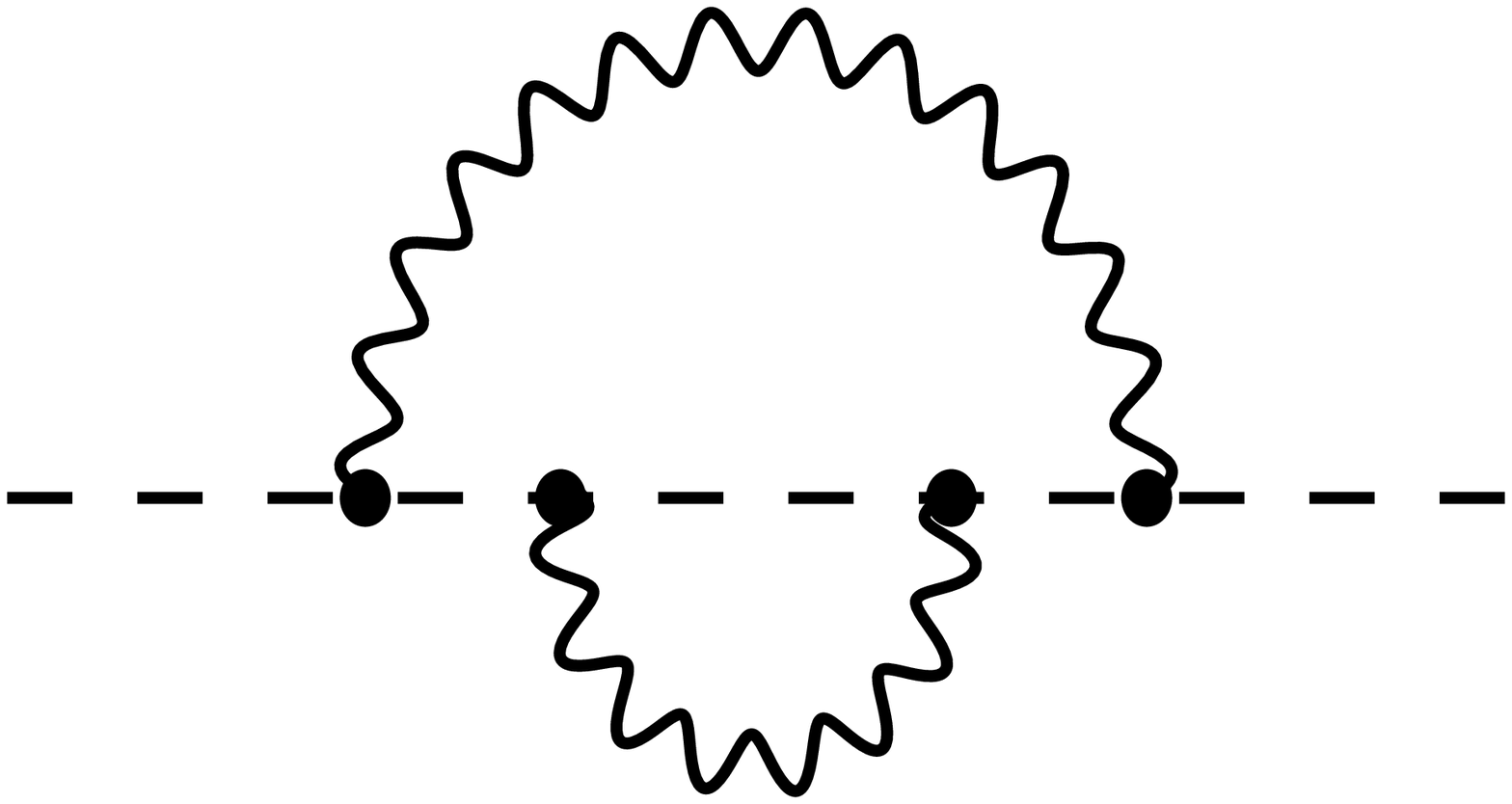}}
		\put(13,6.5){(a4)}

		\put(1,2.6){\includegraphics[scale=0.14]{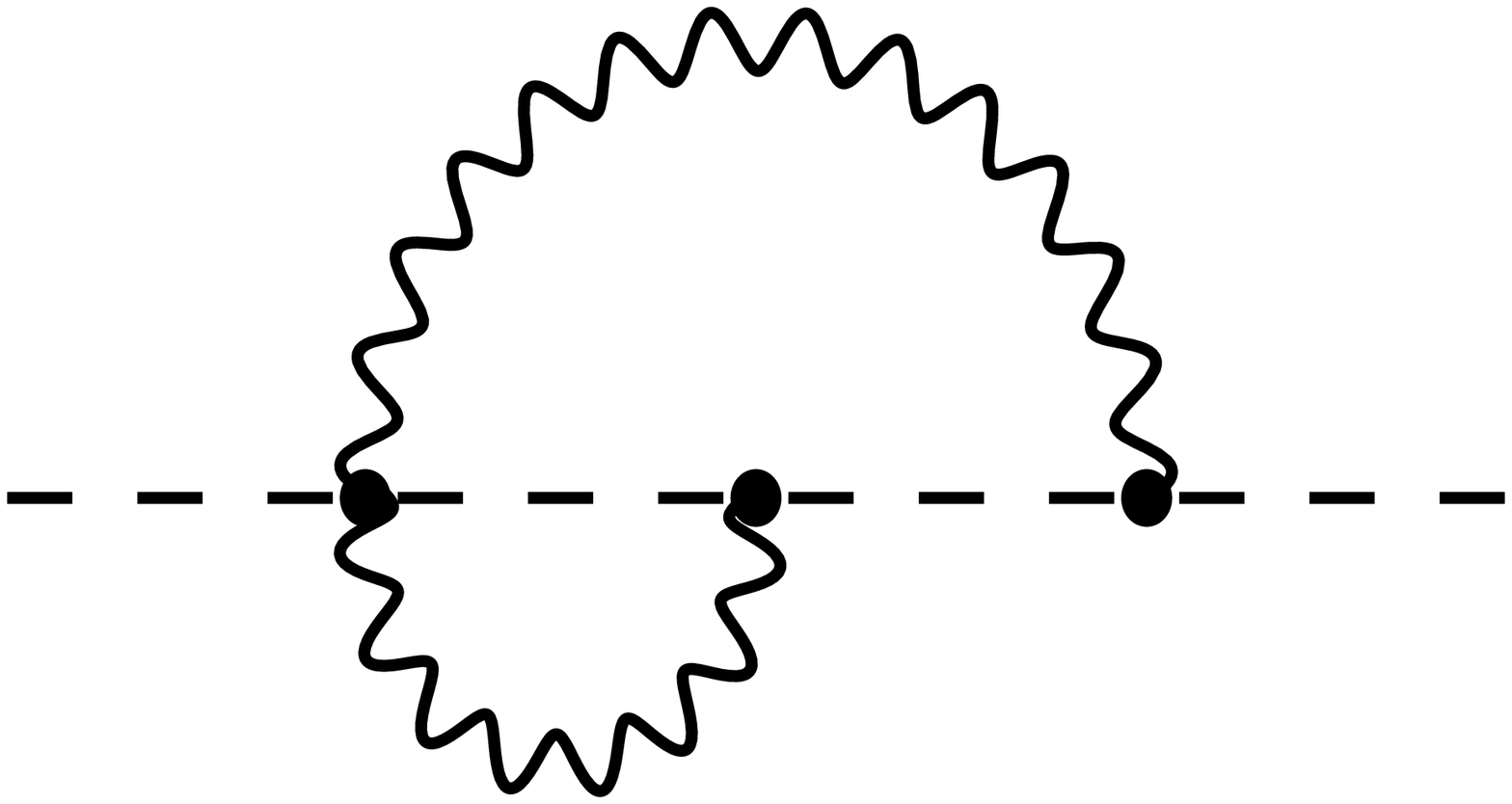}}
		\put(1,4.0){(a5)}
		\put(5,2.6){\includegraphics[scale=0.14]{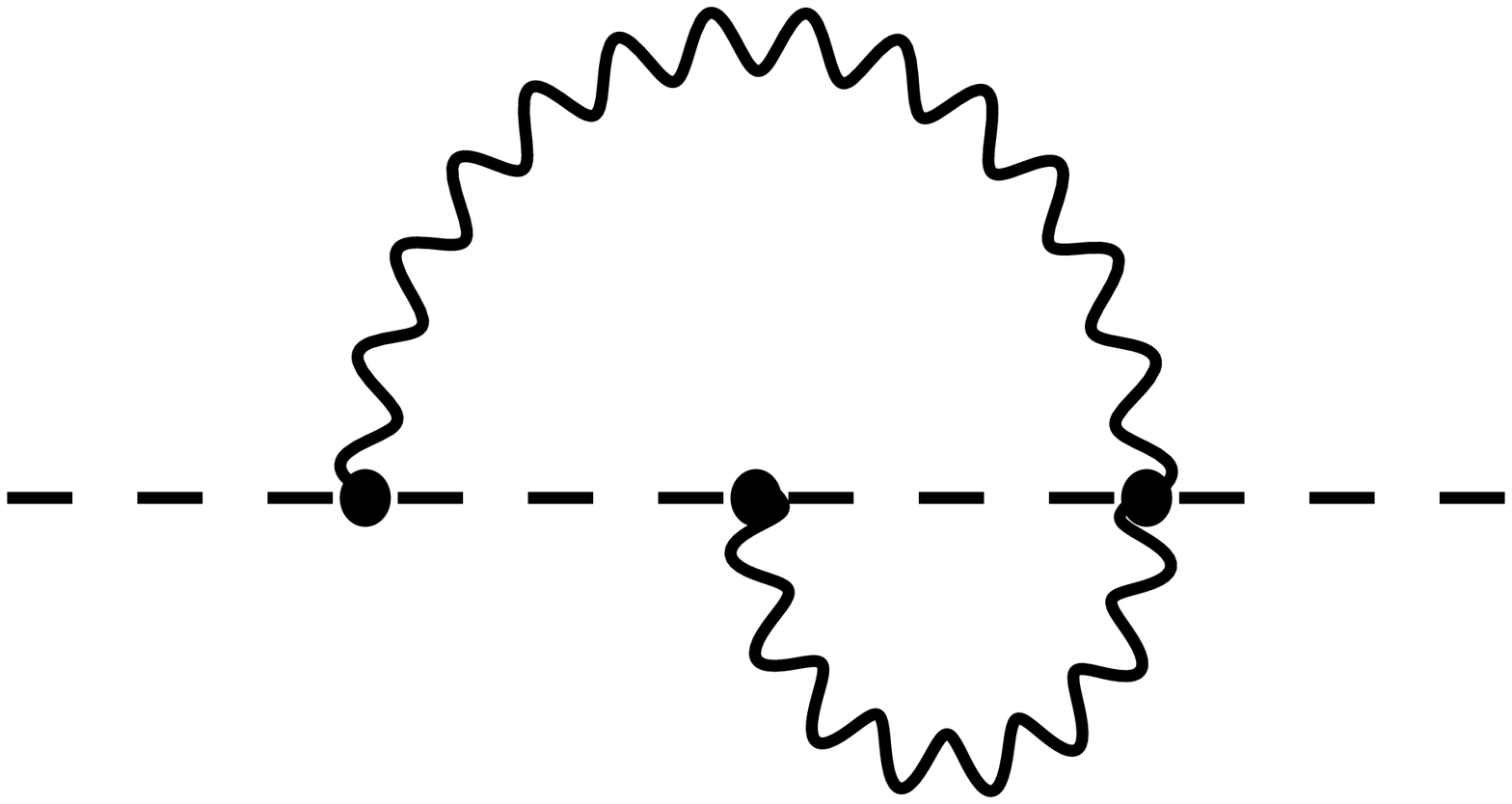}}
		\put(5,4.0){(a6)}
		\put(9,2.6){\includegraphics[scale=0.14]{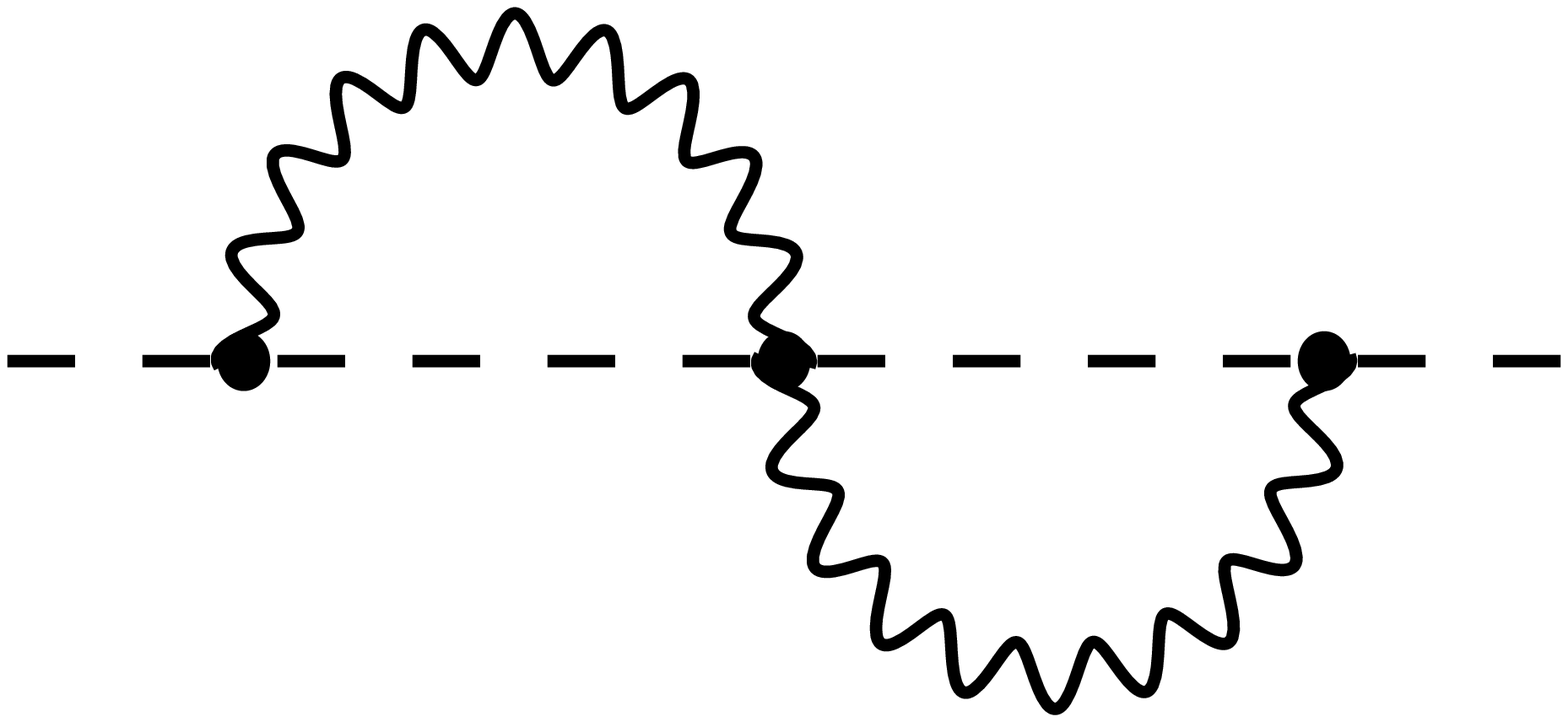}}
		\put(9,4.0){(a7)}
		\put(13,2.5){\includegraphics[scale=0.14]{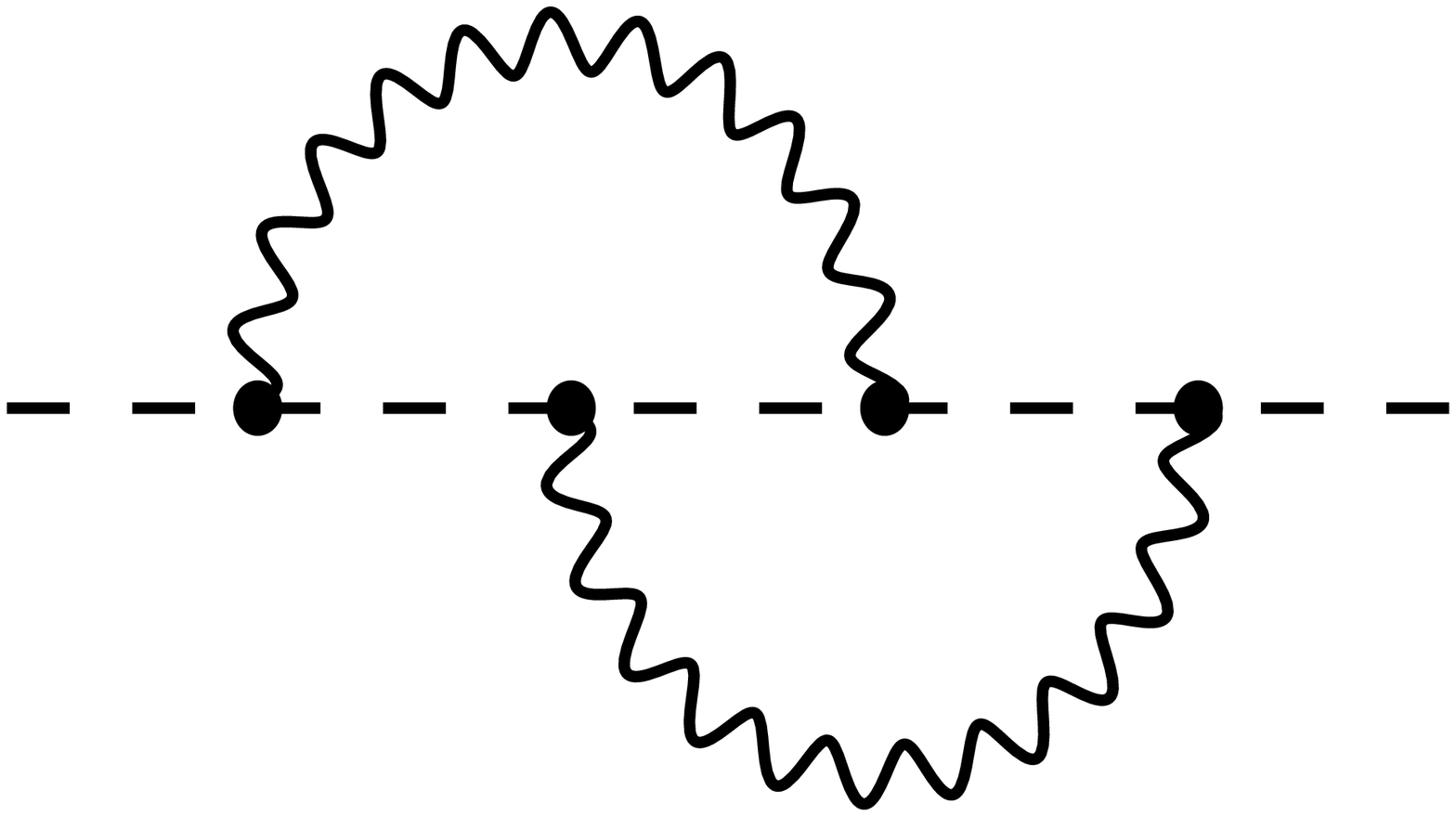}}
		\put(12.8,4.0){(a8)}
		
		\put(3,0){\includegraphics[scale=0.14]{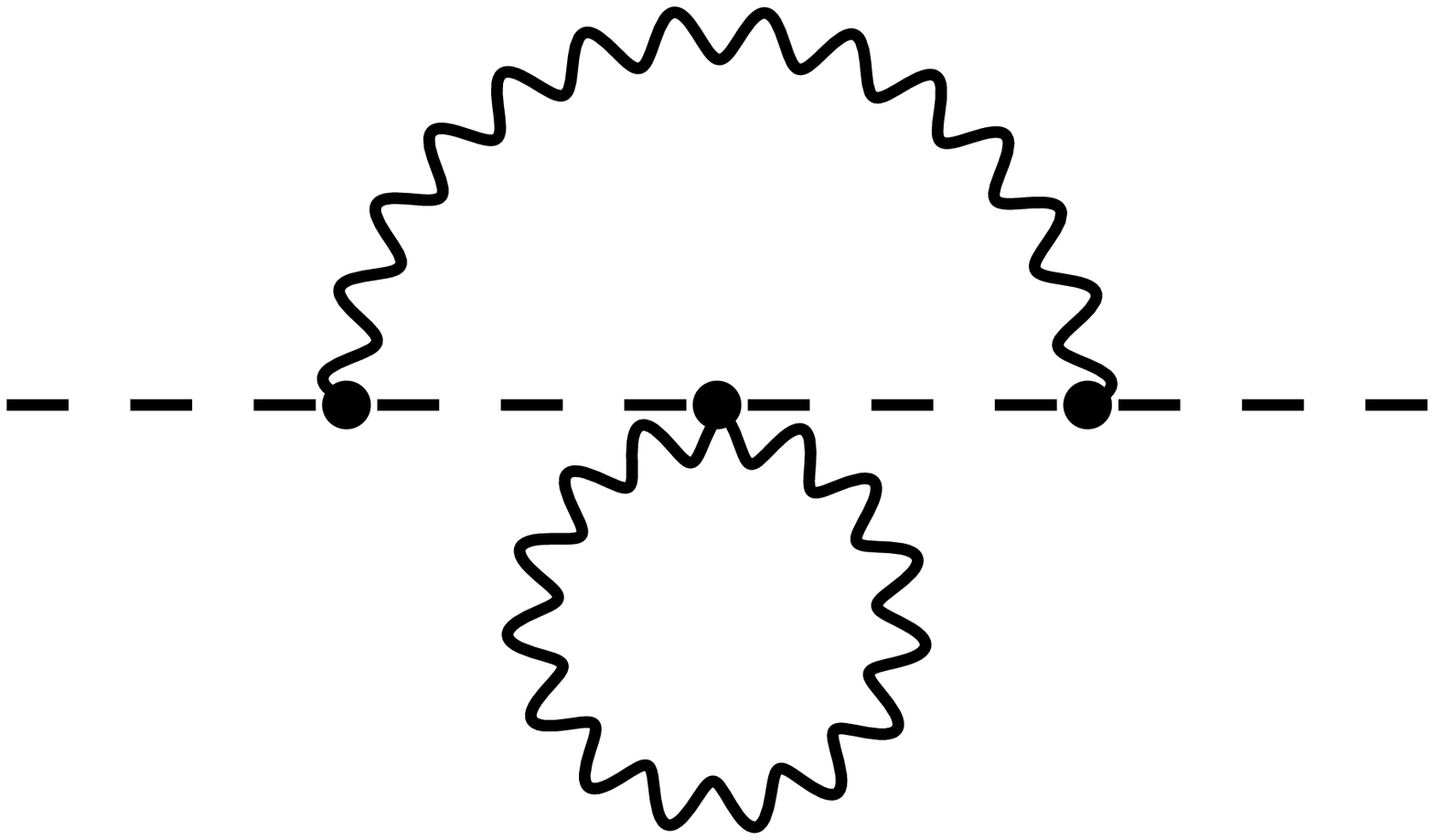}}
		\put(3,1.6){(a9)}
		\put(7.2,0.17){\includegraphics[scale=0.14]{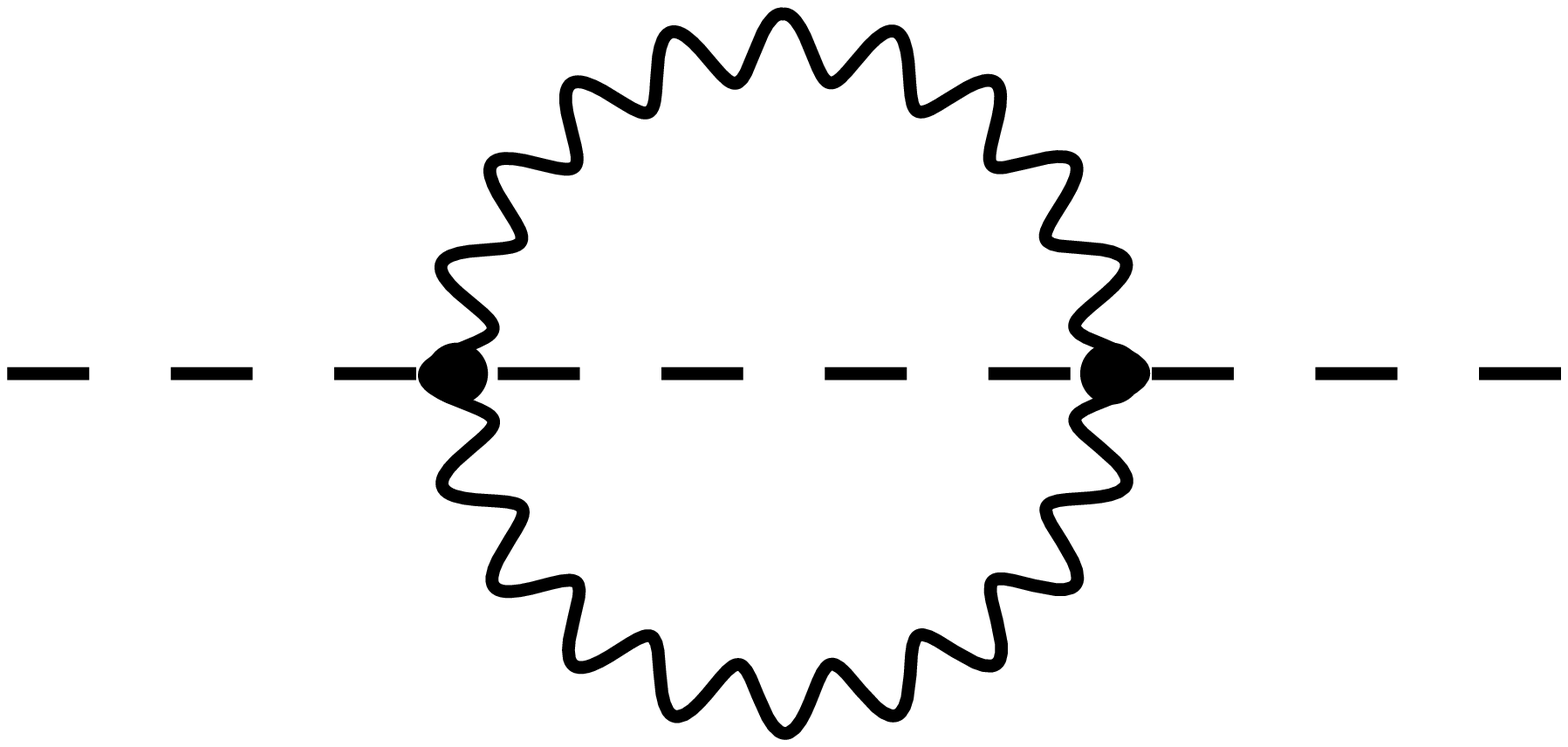}}
		\put(7.2,1.6){(a10)}
		\put(11.3,-0.3){\includegraphics[scale=0.14]{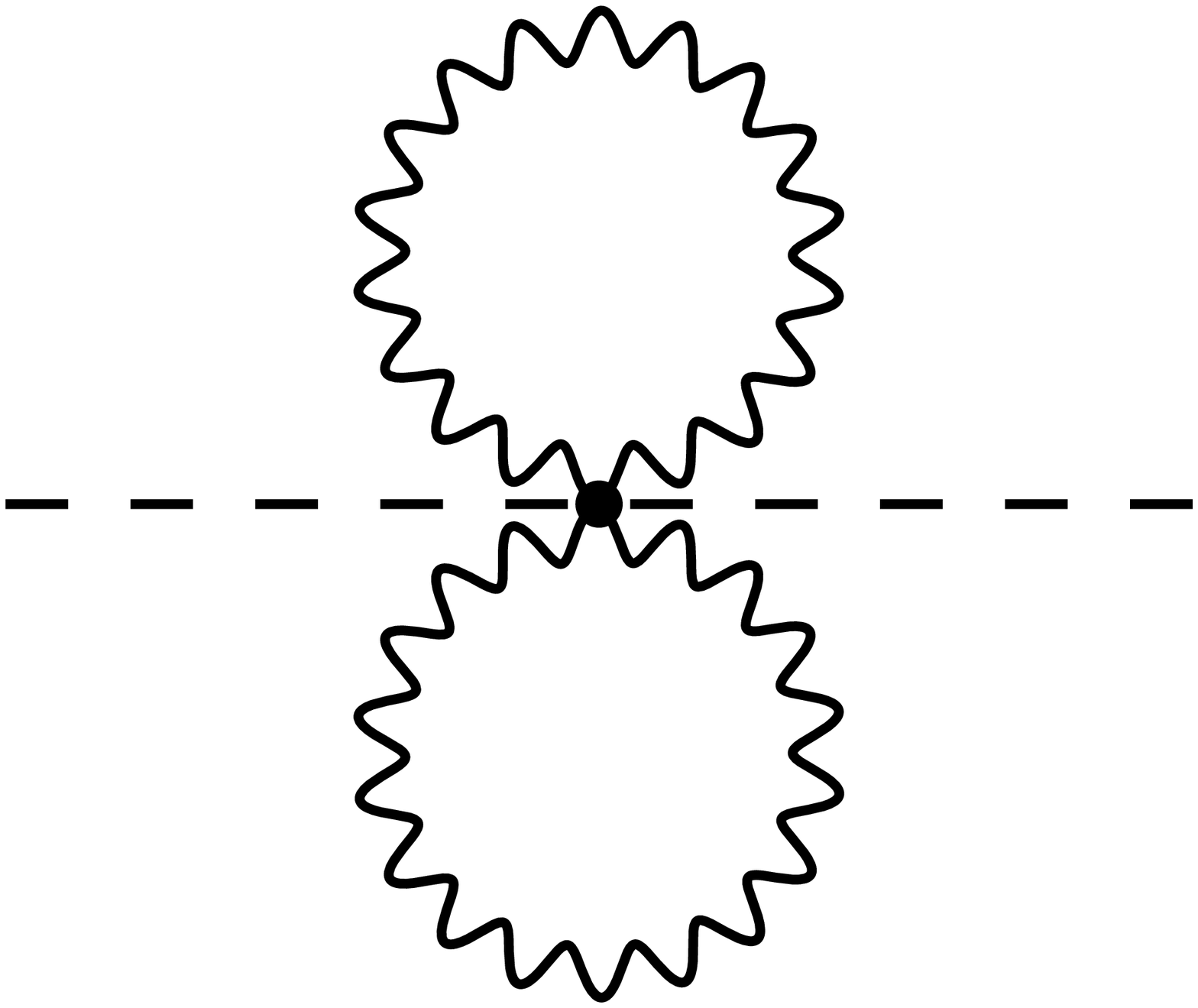}}
		\put(11,1.6){(a11)}
	\end{picture}
\vspace*{3mm}
\caption{Some superdiagrams contributing to the triple gauge-ghost vertices can be obtained by attaching an external gauge line to these supergraphs in all possible ways. In our notation wavy lines correspond to the quantum gauge superfield, while dashed lines denote the Faddeev--Popov ghost propagators and external legs.}
\label{Figure:1:Two-points_a}
\end{figure}

1. The (three-point) superdiagrams of the first group are obtained by attaching an external gauge line to the two-point supergraphs (a1) --- (a11) presented in Fig.~\ref{Figure:1:Two-points_a} in all possible ways.
For instance, the supergraph (a1) produces 9 different three-point superdiagrams contributing to the considered vertex. Four of them arise after attaching an external line to the vertices, while the remaining ones appear when an external line is attached to the propagators. Since a number of superdiagrams obtained from the supergraphs (a1) --- (a11) is rather large, we will not depict them. Note that in the superdiagrams of the first group nonlinear terms inside the function $\mathcal{F}(V)$ are not essential, because they do not produce $O(e_0^5)$ contributions to the effective action which are calculated in the considered approximation. (The nonlinear terms are essential in the one-loop superdiagrams. They will be analyzed below.) Therefore, for constructing the vertices in the considered superdiagrams we can set $\mathcal{F}(V) \to V$ and use the equations

\begin{eqnarray}
	&& \frac{V}{1-e^{2V}} = -\frac{1}{2} + \frac{1}{2} V - \frac{1}{6} V^2 + \frac{1}{90} V^4 + O(V^6);\qquad\\
	&& \frac{V}{1-e^{-2V}} = \frac{1}{2} + \frac{1}{2} V + \frac{1}{6} V^2 - \frac{1}{90} V^4 + O(V^6).
\end{eqnarray}

\noindent
Then from the ghost action (\ref{Action_Faddeev-Popov_Ghosts}) we obtain the interaction terms

\begin{eqnarray}\label{Vertex:gauge-ghost_without_NLR}
	&&\hspace*{-7mm}  \int d^4x\,d^4\theta\,  \Big(\frac{ie_0}{4} f^{ABC} (\bar c^A + \bar c^{+A}) V^B (c^C + c^{+C}) - \frac{e_0^2}{12} f^{ABC}f^{CDE} (\bar c^A + \bar c^{+A}) V^B V^D (c^E - c^{+E})
	\nonumber\\
	&&\hspace*{-7mm} - \frac{e_0^4}{180} f^{ABC} f^{CDE} f^{EFG} f^{GHI} (\bar c^A + \bar c^{+A}) V^B V^D V^F V^H (c^I - c^{+I}) + \ldots \Big).
\end{eqnarray}

\noindent
Note that there are no vertices with an odd number of the quantum gauge superfield lines which is more than 1. (Nevertheless, such vertices can appear if nonlinear terms in the function $\mathcal{F}(V)$ are taken into account.)

Let (A$i$) denotes a set of the superdiagrams constructed by attaching an external gauge line to the supergraph (a$i$) in all possible ways. After the substitution (\ref{Substitution_V}) this external gauge line will correspond to $\bar D^2 H$, and only the contributions to the function $\mathcal{S}$ can be found in this case. The contributions of all sets (A$i$) to this function in the limit of the vanishing external momenta (obtained after a rather complicated calculation) are given by the expressions

\begin{eqnarray}\label{result_A_first}
	&& \hspace*{-8mm} \Delta_{\mbox{\scriptsize A1}}\mathcal{S} = \frac{e_0^4}{24} (C_2)^2 \int \frac{d^4K}{(2\pi)^4} \frac{d^4L}{(2\pi)^4} \frac{\xi_0^2}{K_K K^4 K_L L^2 } \bigg( \frac{2}{(K+L)^2} - \frac{1}{L^2}  \bigg);\\
	&& \hspace*{-8mm} \Delta_{\mbox{\scriptsize A2}} \mathcal{S} = \frac{e_0^4}{72} (C_2)^2 \int \frac{d^4K}{(2\pi)^4} \frac{d^4L}{(2\pi)^4} \frac{\xi_0}{K_K K^2 L^2} \bigg( \Big( \frac{\xi_0}{K_L}-\frac{1}{R_L} \Big) \frac{1}{L^2} \Big(\frac{1}{K^2}
	 - \frac{1}{(K+L)^2}  \Big)
	 \nonumber\\
	 && \hspace*{-8mm}  - \Big( \frac{\xi_0}{K_L}+\frac{1}{R_L} \Big) \frac{1}{K^2 (K+L)^2} \bigg);\\
	 && \hspace*{-8mm} \Delta_{\mbox{\scriptsize A4}} \mathcal{S} = \frac{e_0^4}{8} (C_2)^2 \int \frac{d^4K}{(2\pi)^4} \frac{d^4L}{(2\pi)^4} \frac{\xi_0}{K_K K^2 L^2 } \Big( \frac{\xi_0}{K_L}-\frac{1}{R_L}  \Big) \bigg(  \frac{1}{L^2K^2} + \frac{1}{(K+L)^2} \Big(\frac{1}{L^2} - \frac{1}{K^2}  \Big)  \bigg);\nonumber\\
    &&\vphantom{1}\\
	 && \hspace*{-8mm} \Delta_{\mbox{\scriptsize A5}} \mathcal{S} = - \frac{e_0^4}{9} (C_2)^2 \int \frac{d^4K}{(2\pi)^4} \frac{d^4L}{(2\pi)^4} \frac{\xi_0}{K_K K^2 L^2 } \bigg( \Big( \frac{\xi_0}{4 K_L}+\frac{1}{R_L}  \Big) \frac{1}{K^2 (K+L)^2} + \Big( \frac{\xi_0}{K_L}-\frac{1}{R_L}  \Big)
	 \nonumber\\
	 && \hspace*{-8mm} \times \frac{1}{L^2}\Big(\frac{1}{(K+L)^2} + \frac{1}{K^2}   \Big)    \bigg); \\
	 && \hspace*{-8mm} \Delta_{\mbox{\scriptsize A8}} \mathcal{S} = \frac{e_0^4}{24} (C_2)^2 \int \frac{d^4K}{(2\pi)^4} \frac{d^4L}{(2\pi)^4} \frac{\xi_0^2}{K_K K^4 K_L L^2 } \bigg( \frac{2}{(K+L)^2} + \frac{1}{L^2}  \bigg);\\
	 && \hspace*{-8mm} \Delta_{\mbox{\scriptsize A9}} \mathcal{S} = - \frac{e_0^4}{6} (C_2)^2 \int \frac{d^4K}{(2\pi)^4} \frac{d^4L}{(2\pi)^4} \frac{\xi_0}{K_K K^4 L^4 } \Big( \frac{\xi_0}{K_L}-\frac{1}{R_L}  \Big);\\\nonumber\\
	 && \hspace*{-8mm} \Delta_{\mbox{\scriptsize A3}} \mathcal{S} = \Delta_{\mbox{\scriptsize A6}} \mathcal{S} = \Delta_{\mbox{\scriptsize A7}} \mathcal{S} = \Delta_{\mbox{\scriptsize A10}} \mathcal{S} = \Delta_{\mbox{\scriptsize A11}} \mathcal{S} = 0.
\end{eqnarray}

\noindent
They are written after the Wick rotation, and the Euclidean momenta are denoted by capital letters. For the regulator functions in the momentum representation we use the notations $K_K\equiv K(K^2/\Lambda^2)$, $R_K\equiv R(K^2/\Lambda^2)$ etc.

\begin{figure}[h]
	\begin{picture}(0,3.4)		
		\put(2,0.4){\includegraphics[scale=0.26,clip]{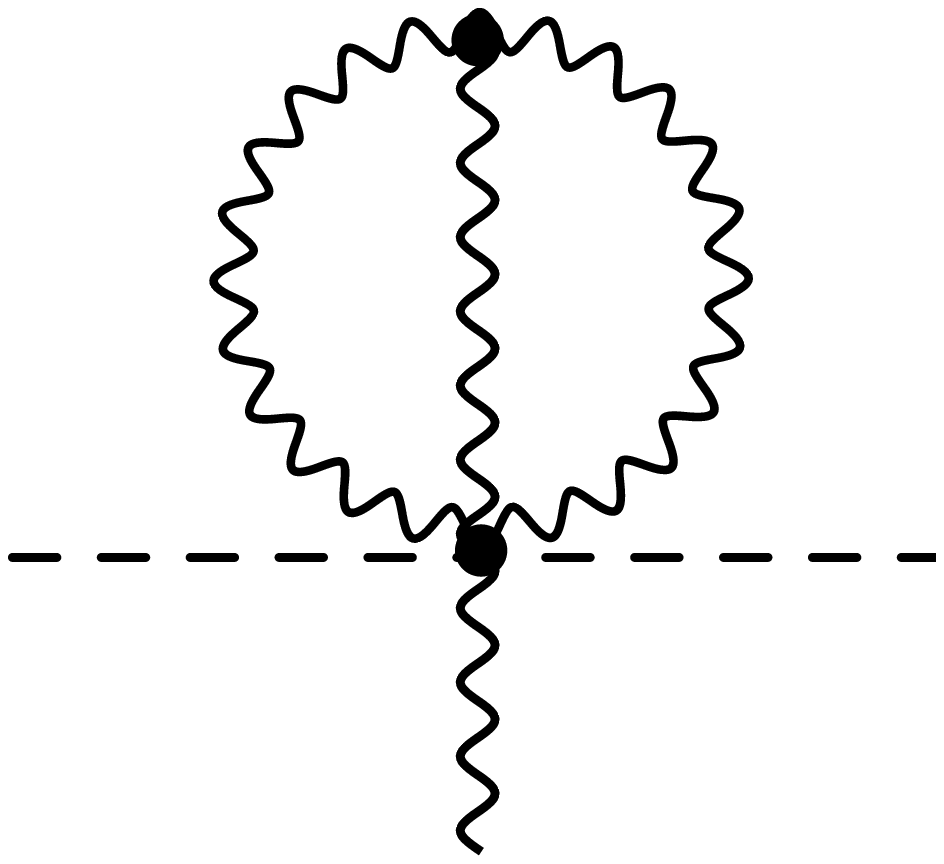}}
		\put(2,2.8){(b1)}
		\put(2,0.8){$ \bar{c}^+ $}
		\put(4.4,0.9){$ c$}
		\put(2.3,0.3){$ \bar{D}^2 H$}

		\put(6.5,0.3){\includegraphics[scale=0.3,clip]{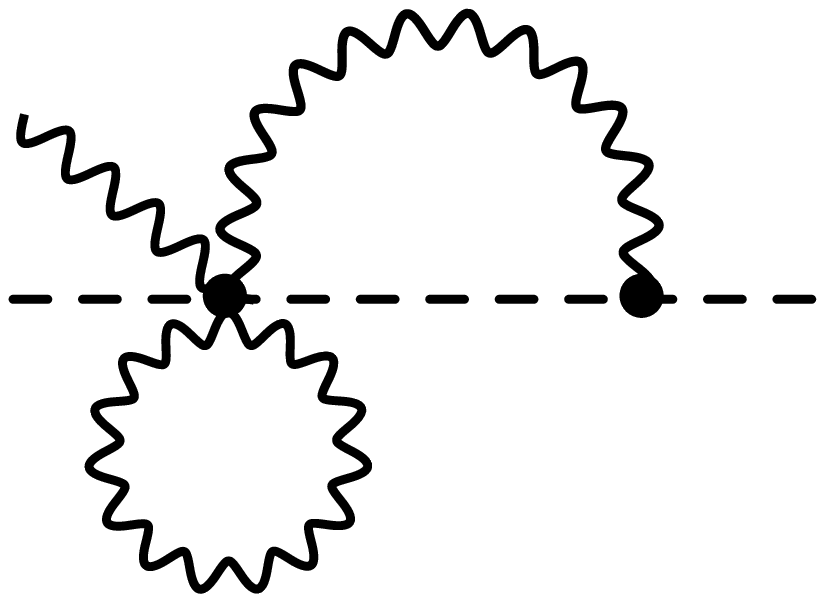}}
		\put(6.5,2.8){(b2)}
		
		\put(10.5,0.3){\includegraphics[scale=0.3,clip]{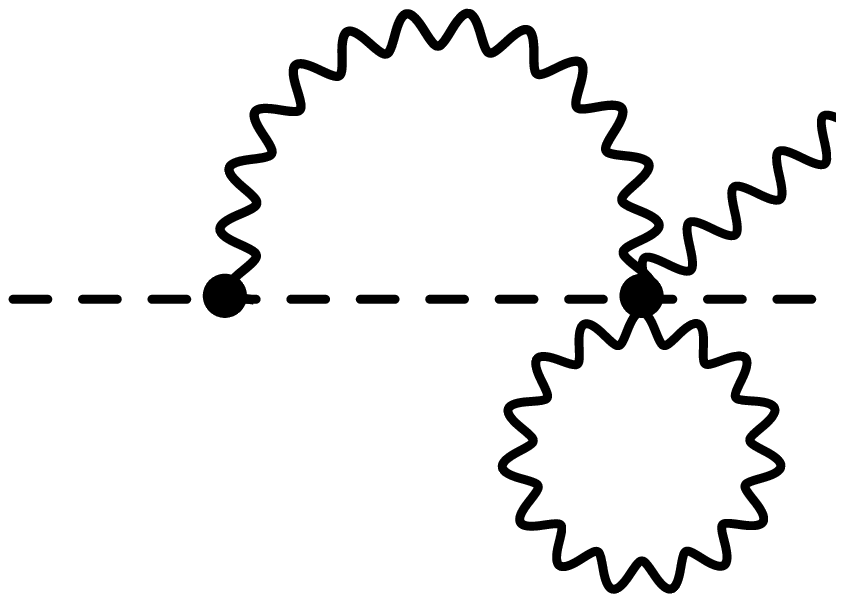}}
		\put(10.5,2.8){(b3)}		
	\end{picture}
\caption{It is convenient to analyze these three-point superdiagrams together with the ones constructed from the superdiagrams in Fig.~\ref{Figure:1:Two-points_a} as explained in the text.}
\label{Figure:2:Three-points_2-loops_B}
\end{figure}

2. The second group of superdiagrams contributing to the considered vertex is presented in Fig.~\ref{Figure:2:Three-points_2-loops_B}. Their contributions to the function $\mathcal{S}$ in the limit of the vanishing external momenta are

\begin{eqnarray}
	&& \Delta_{\mbox{\scriptsize b2}} \mathcal{S} = \frac{5 e_0^4}{36}  (C_2)^2 \int \frac{d^4K}{(2\pi)^4} \frac{d^4L}{(2\pi)^4} \frac{\xi_0}{K_K K^4 L^4 } \Big( \frac{\xi_0}{K_L}-\frac{1}{R_L}  \Big);\\ \nonumber\\
	&&\Delta_{\mbox{\scriptsize b1}} \mathcal{S} = \Delta_{\mbox{\scriptsize b3}} \mathcal{S} = 0. \label{result_b_last}
\end{eqnarray}

Summing up the expressions (\ref{result_A_first}) --- (\ref{result_b_last}), we see that the total contribution of the first and second groups of the superdiagrams to the function $\mathcal{S}$ vanishes in the limit of the vanishing external momenta,

\begin{eqnarray}
	\sum_{i=1}^{11} \Delta_{\mbox{\scriptsize A}i} \mathcal{S} + \sum_{i=1}^{3}\Delta_{\mbox{\scriptsize b}i} \mathcal{S} = 0.
\end{eqnarray}

\begin{figure}[h]
	\begin{picture}(0,4.8)
		\put(3.1,3.1){\includegraphics[scale=0.18,clip]{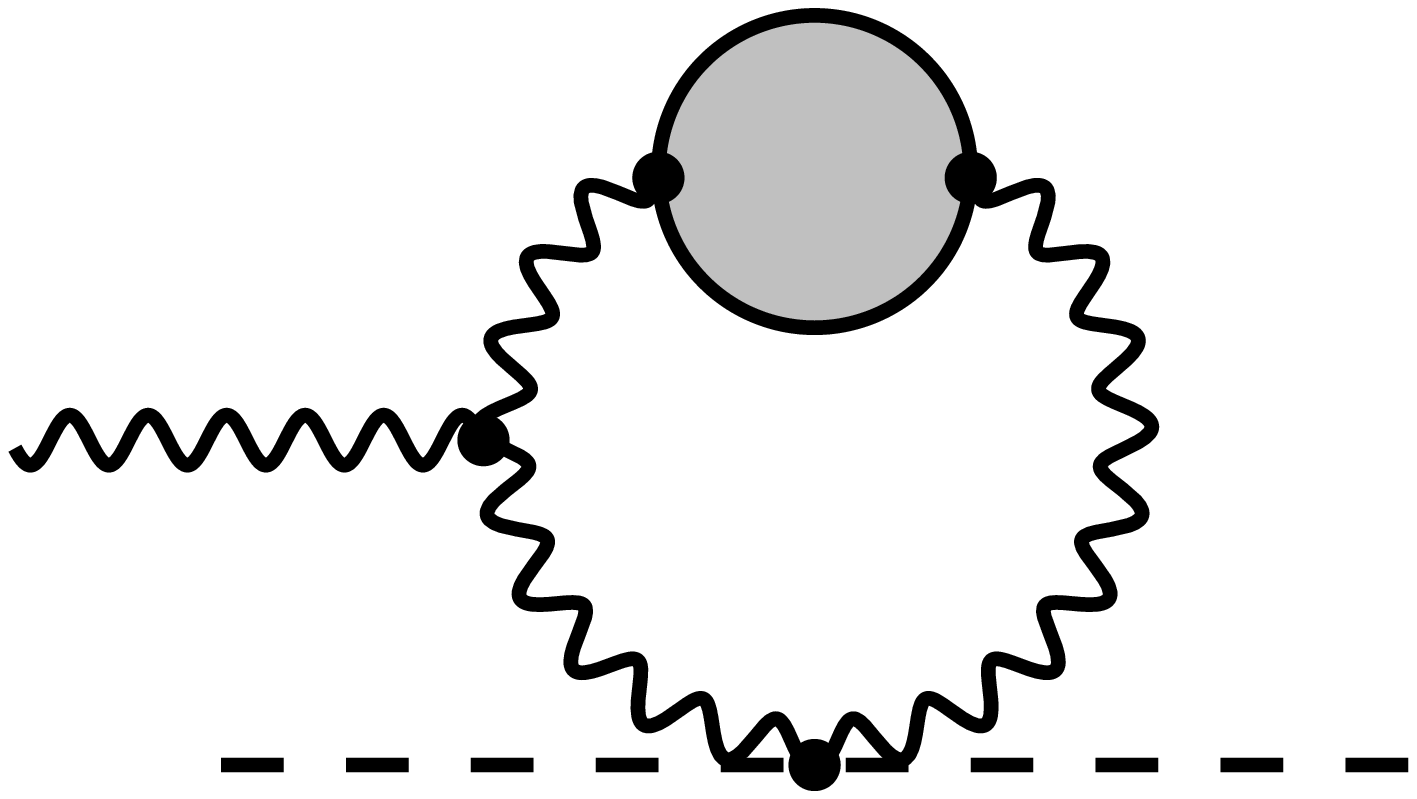}}
		\put(3.5,4.7){(c1)}
		\put(3.2,2.7){$ \bar{c}^+ $}
		\put(5.6,2.7){$ c$}
		\put(2.7,3.9){$ \bar{D}^2 H$}
		
		\put(7.2,3.1){\includegraphics[scale=0.2,clip]{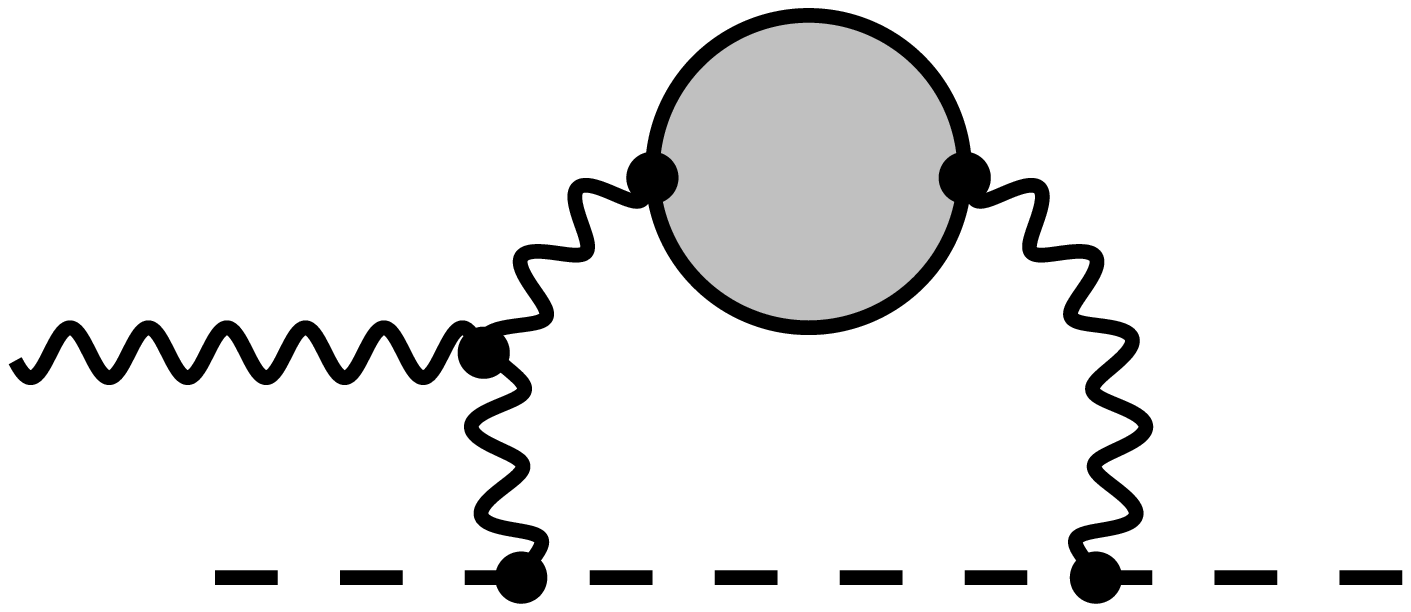}}
		\put(7.5,4.7){(c2)}
		\put(11.5,3.1){\includegraphics[scale=0.2,clip]{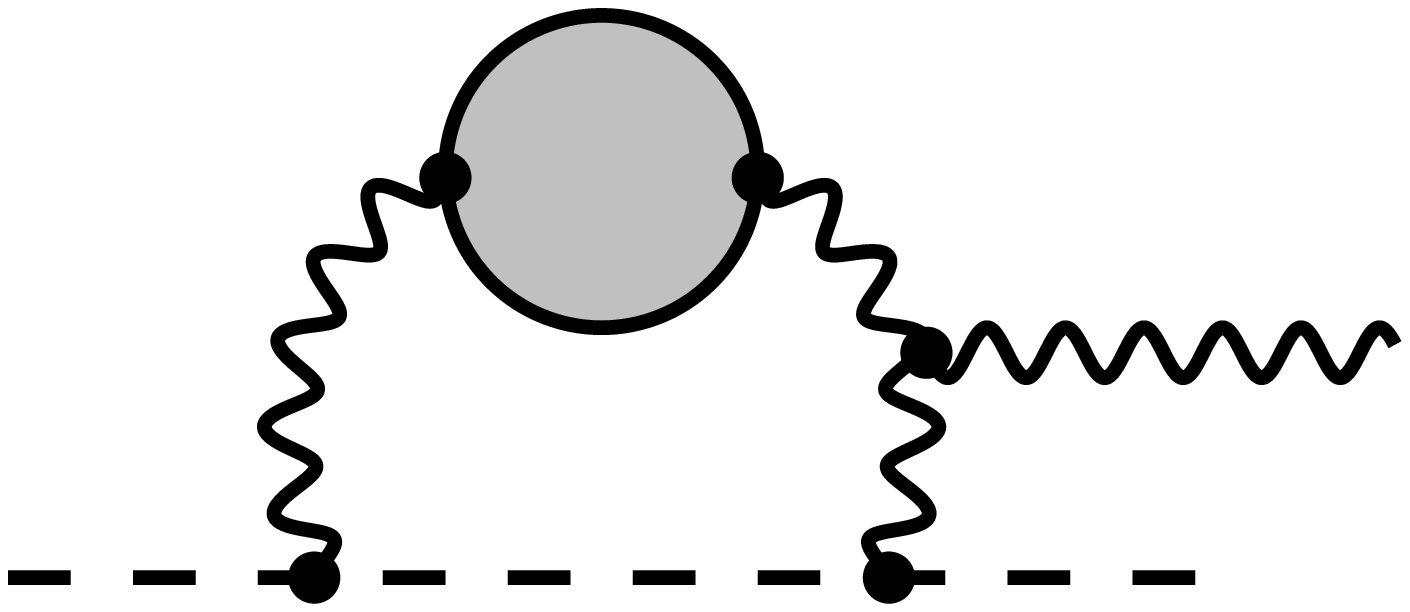}}
		\put(11.5,4.7){(c3)}

		\put(3.5,-0.1){\includegraphics[scale=0.2,clip]{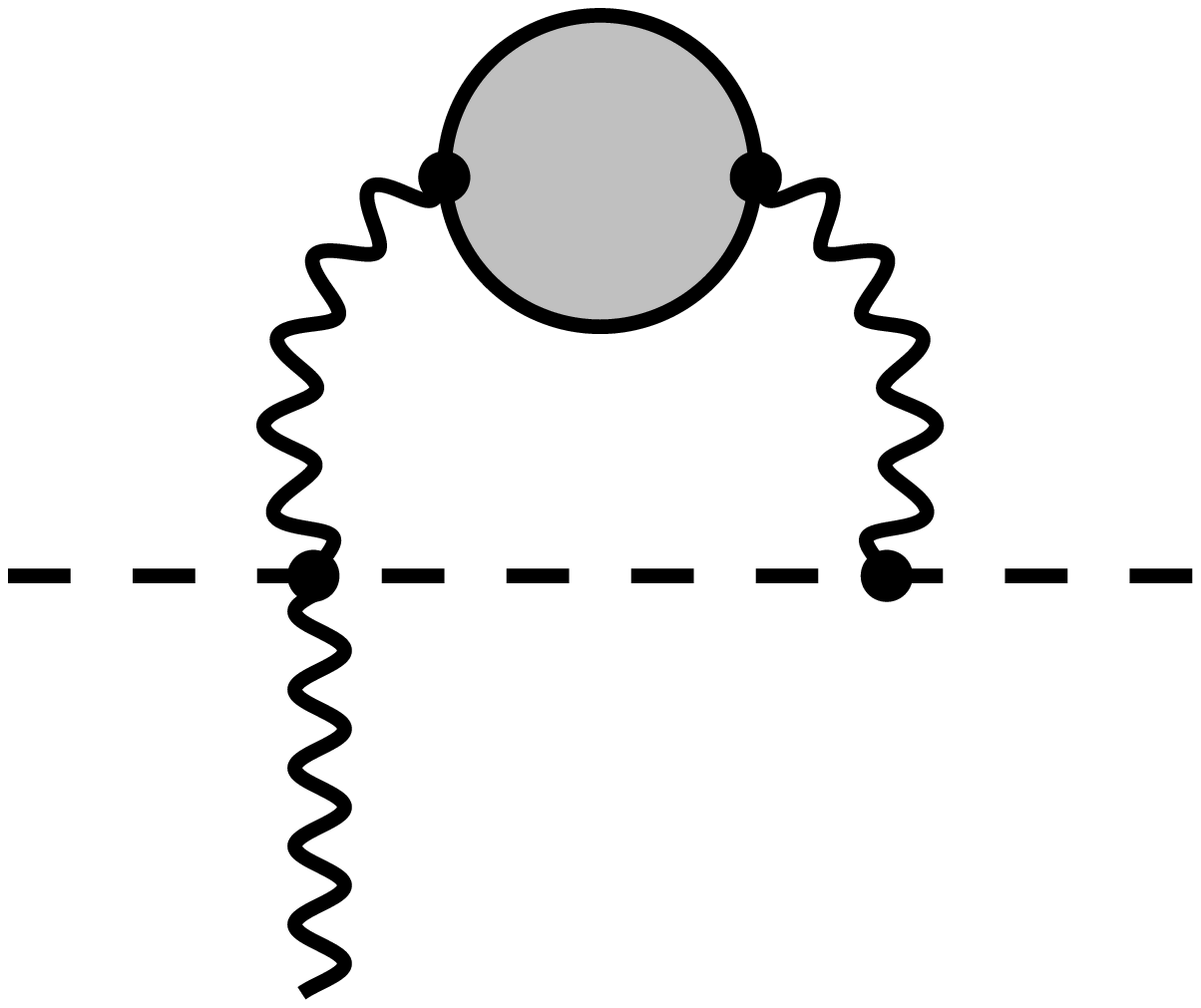}}
		\put(3.5,2.1){(c4)}
		\put(7.5,-0.1){\includegraphics[scale=0.2,clip]{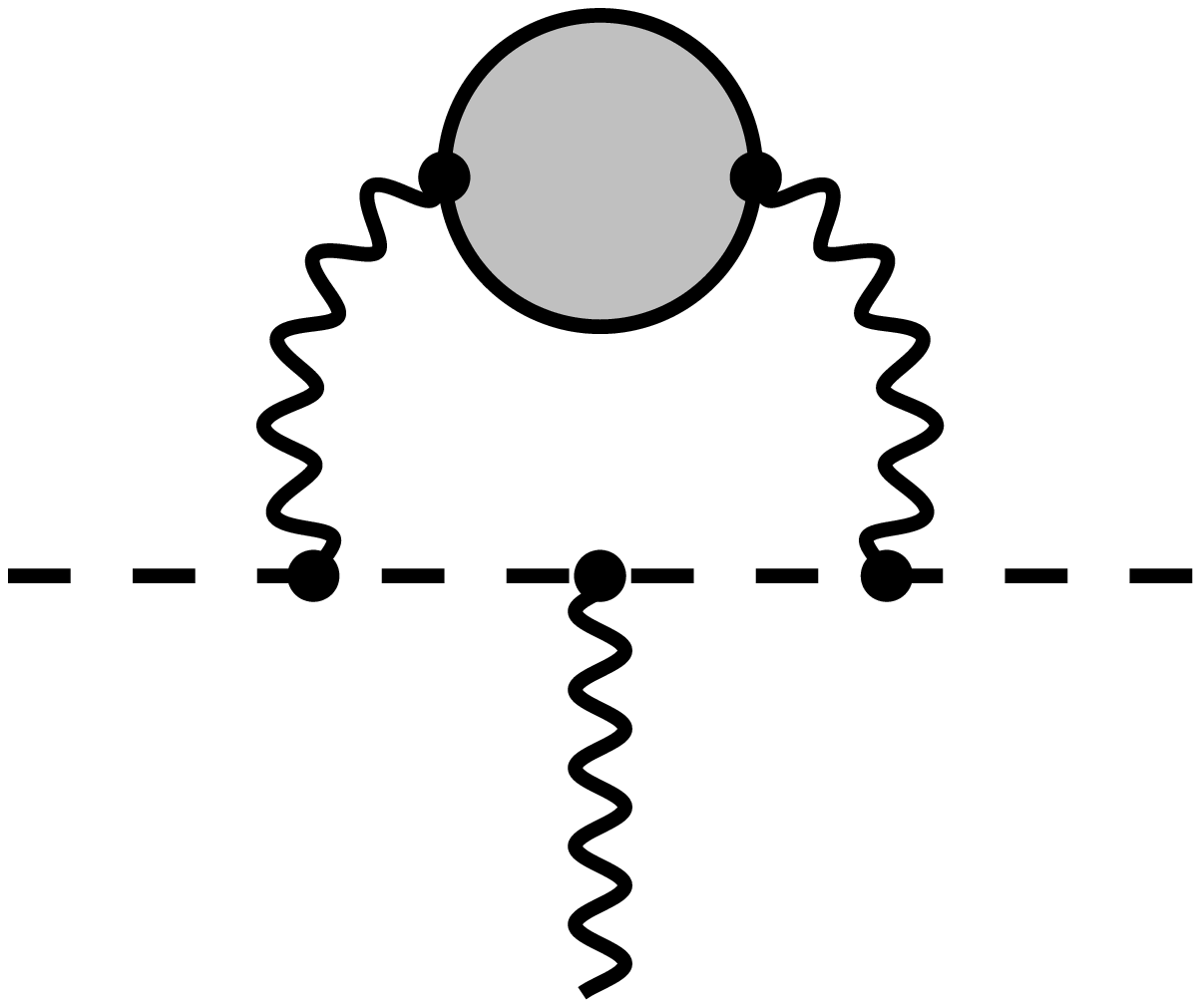}}
		\put(7.5,2.1){(c5)}
		\put(11.5,-0.1){\includegraphics[scale=0.2,clip]{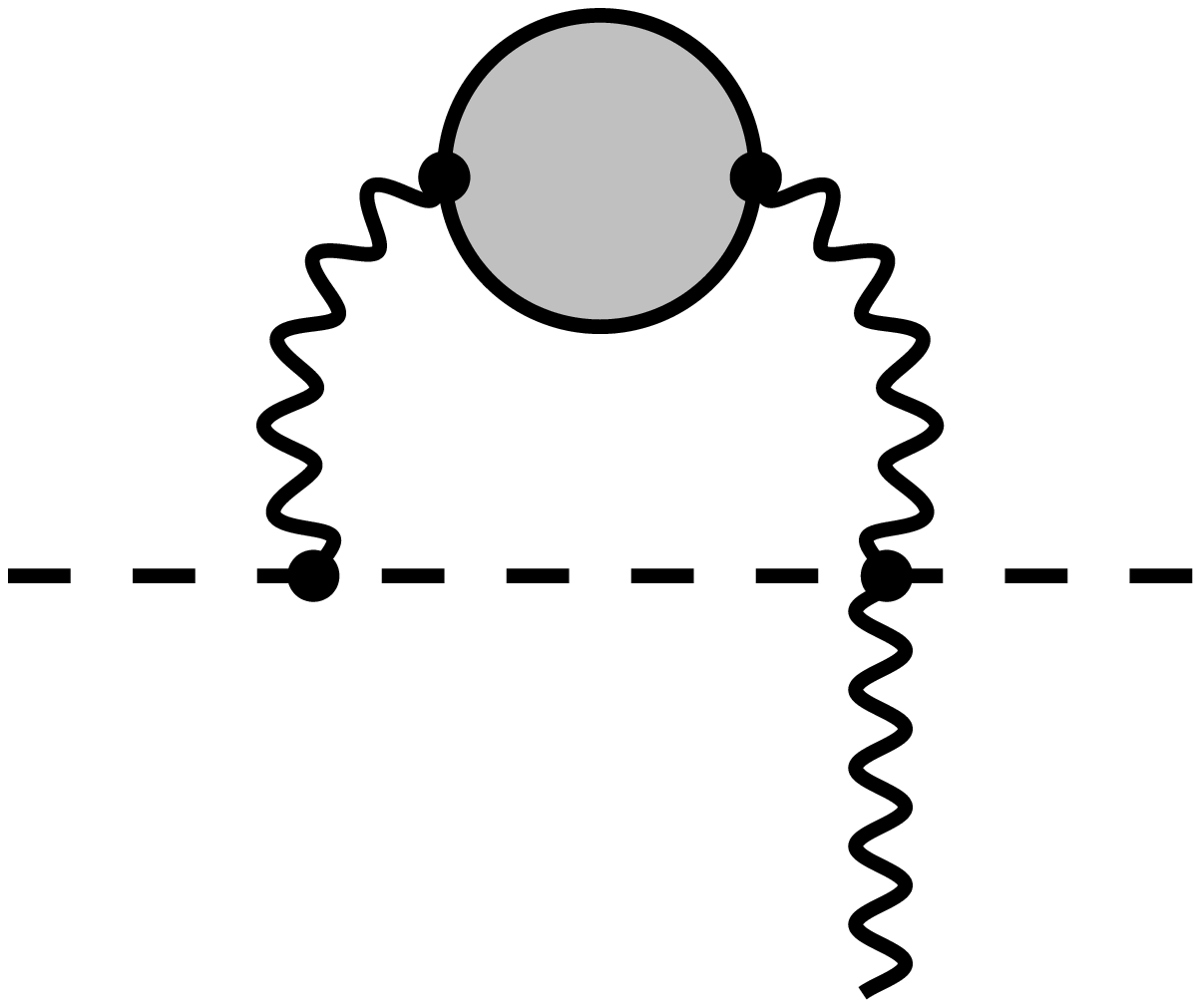}}
		\put(11.5,2.1){(c6)}	
	\end{picture}
\caption{Two-loop superdiagrams containing an insertion of the one-loop quantum gauge superfield polarization operator denoted by a gray disk.}
\label{Figure:3:Three-points_2-loops_Polarization_C}
\end{figure}

\begin{figure}[b!]
	\begin{picture}(0,3.5)
		\put(0.5,2.4){\includegraphics[scale=0.25,clip]{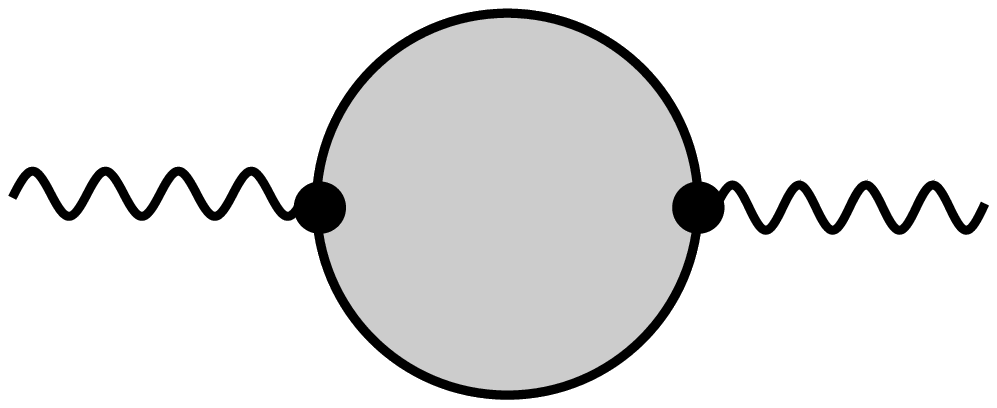}}
		
		\put(2.8,3.1){$ V $}
		\put(0.5,3.1){$ V $}
		
		\put(3.5,2.8){$ \to $}
		
		\put(4.4,2.3){\includegraphics[scale=0.25,clip]{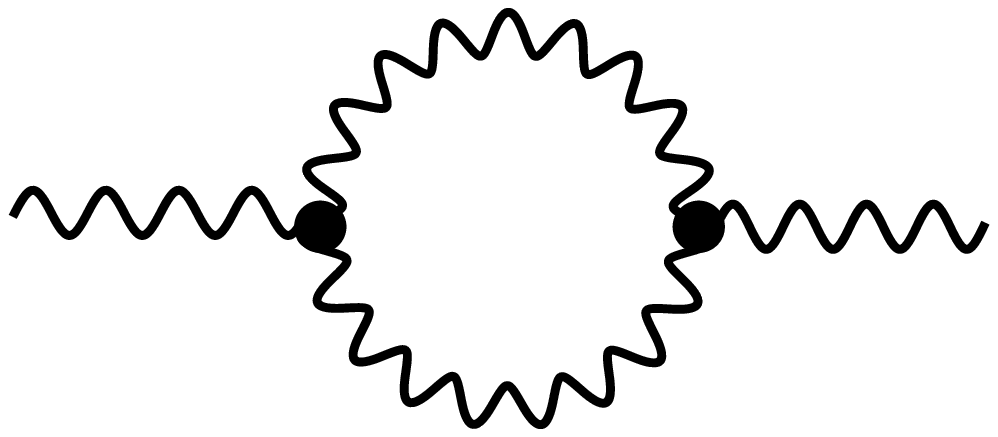}}
		\put(7.2,2.7){$ + $}
		
		\put(8,2.35){\includegraphics[scale=0.25,clip]{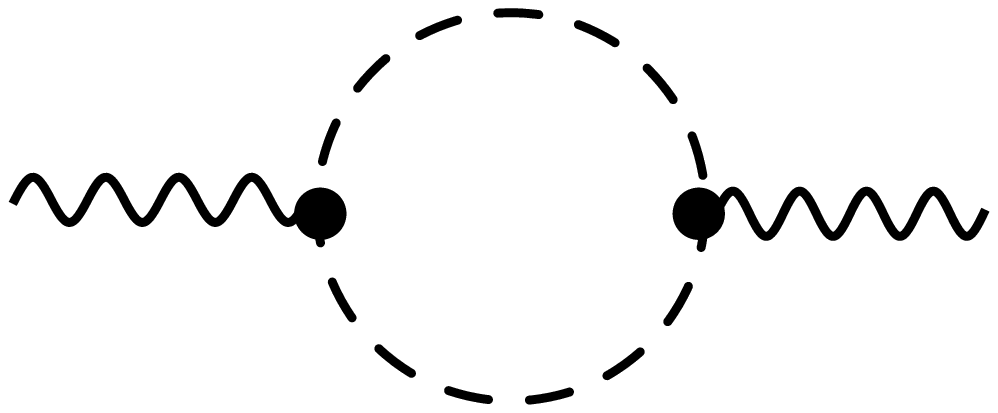}}
		\put(10.9,2.7){$ + $}
		
		\put(11.8,2.35){\includegraphics[scale=0.25,clip]{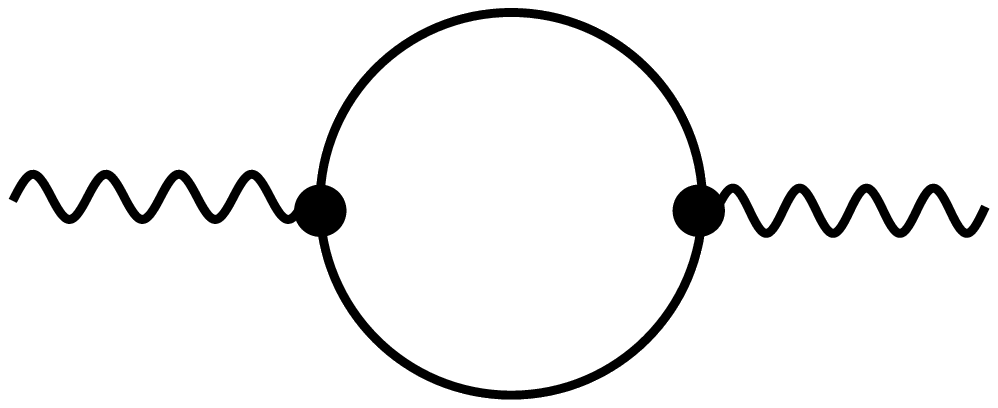}}
		
		\put(3.5,0.8){$ + $}
		\put(4.8,0){\includegraphics[scale=0.23,clip]{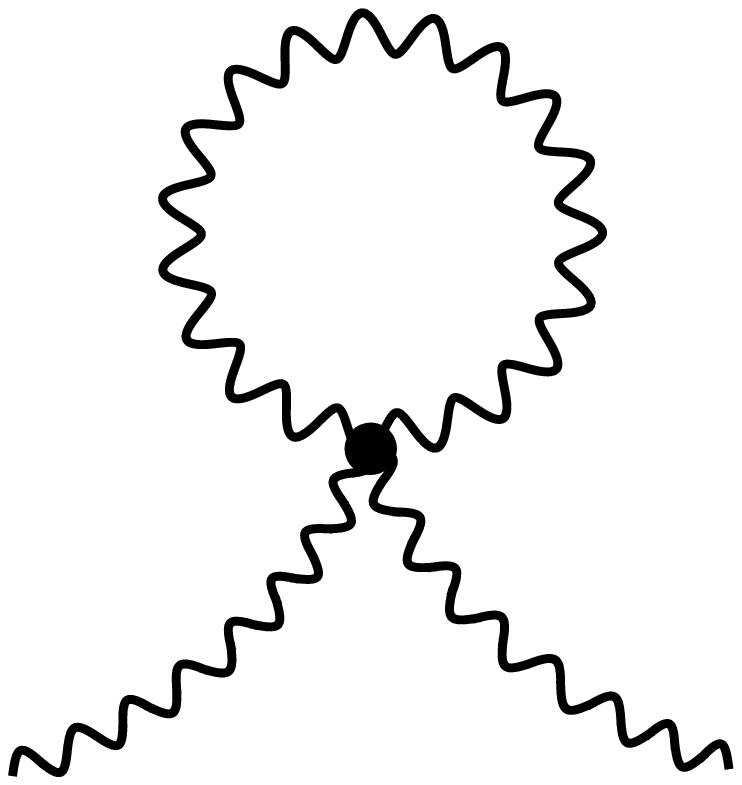}}
		
		\put(7.2,0.8){$ + $}
		\put(8.45,0){\includegraphics[scale=0.23,clip]{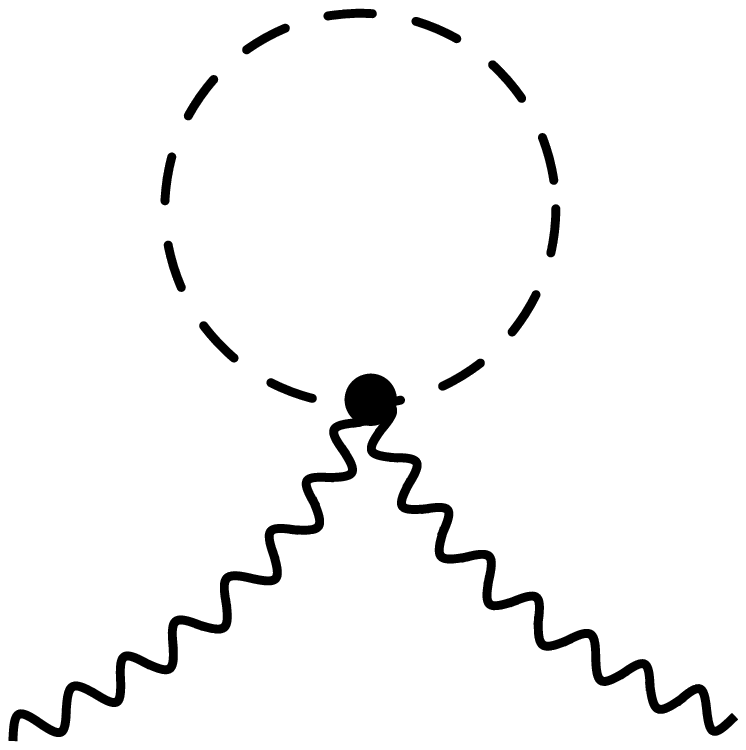}}
		
		\put(10.9,0.8){$ + $}
		\put(12.25,0.05){\includegraphics[scale=0.23,clip]{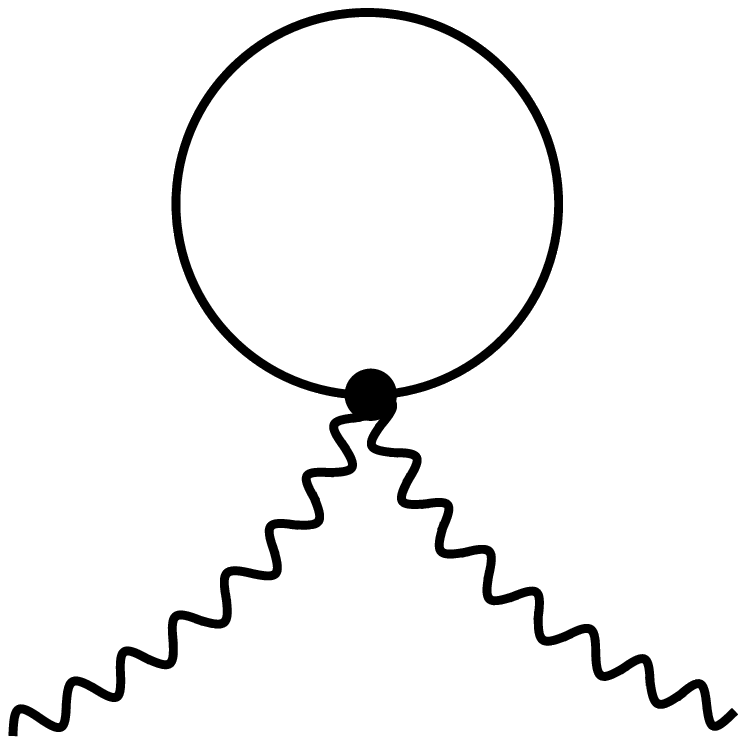}}		
	\end{picture}
\caption{One-loop superdiagrams contributing to the part of the quantum gauge superfield polarization operator proportional to $C_2$. Here the solid lines denote the propagators of the Pauli--Villars superfields $\varphi_{1,2,3}$, while similar supergraphs with the internal lines of the matter superfields and of the Pauli--Villars superfields $\Phi_i$ have already been considered in Ref.~\cite{Kuzmichev:2021yjo}.}
\label{Figure:4:Polarization_Operator_Gauge_Ghosts_PV}
\end{figure}

3. The third group of superdiagrams is depicted in Fig.~\ref{Figure:3:Three-points_2-loops_Polarization_C}. These superdiagrams include an insertion of the polarization operator of the quantum gauge superfield, which we will denote by $\Pi$. In our notation it is defined by the equation

\begin{equation}\label{Polarization_Operator:definition}
    \Pi(\alpha_0,\lambda_0,k^2/\Lambda^2) \equiv R(k^2/\Lambda^2) - \alpha_0 d_q^{-1}(\alpha_0,\lambda_0,k^2/\Lambda^2).
\end{equation}

\noindent
Here the function $d_q^{-1}$ is related to the part of the effective action corresponding to the two-point Green function of the quantum gauge superfield,

\begin{equation}\label{Dq_Definition}
	\Gamma^{(2)}_V - S_{\mbox{\scriptsize gf}}^{(2)} = - \frac{1}{8\pi} \mbox{tr} \int \frac{d^4k}{(2\pi)^4}\, d^4\theta\, V(-k,\theta) \partial^2 \Pi_{1/2} V(k,\theta)\, d_q^{-1}\big(\alpha_0,\lambda_0,k^2/\Lambda^2\big),
\end{equation}

\noindent
which is transversal due to the Slavnov--Taylor identities \cite{Taylor:1971ff,Slavnov:1972fg}.

Partially a contribution of the superdiagrams in Fig.~\ref{Figure:3:Three-points_2-loops_Polarization_C} (which is proportional to $C_2 T(R)$ and comes from the supergraphs containing a loop of the matter superfields or a loop of the Pauli--Villars superfields $\Phi_i$) has already been calculated in Ref.~\cite{Kuzmichev:2021yjo}. In this paper we will be interested in such superdiagrams that contain an insertion of a loop of the quantum gauge superfield, of the Faddeev--Popov ghosts, or of the Pauli--Villars superfields  $\varphi_{1,2,3}$. The subdiagrams contributing to the corresponding part of the polarization operator (denoted by $\Delta\Pi$) are shown in Fig.~\ref{Figure:4:Polarization_Operator_Gauge_Ghosts_PV}. Their sum has been calculated in Ref.~\cite{Kazantsev:2017fdc} and (as a function of the Euclidean momentum) is written as

\begin{eqnarray}\label{Delta_Pi}
	\Delta\Pi (\alpha_0, K^2/\Lambda^2 )= -8\pi\alpha_0 C_2 \Big(f(K/\Lambda) + g(\xi_0, K/\Lambda)\Big),
\end{eqnarray}

\noindent
where analytic expressions for $f(K/\Lambda)$ and $g(\xi_0,K/\Lambda)$ can be found in Ref.~\cite{Kazantsev:2017fdc}. (Note that, as we have already mentioned, the full one-loop expression for the polarization operator also contains a matter loop contribution proportional to $T(R)$, which was not included into Eq. (\ref{Delta_Pi}).) In terms of the polarization operator $\Pi$ the exact propagator of the quantum gauge superfield can be written in the form

\begin{equation}\label{Effective_Propagator_V}
	2i\left(\frac{1}{\big(R -\Pi\big)\partial^2} - \frac{1}{16\partial^4}\Big(D^2 \bar D^2 + \bar D^2 D^2\Big)\Big(\frac{\xi_0}{K} - \frac{1}{R-\Pi}\Big)\right) \delta^8_{xy} \delta^{AB},
\end{equation}

\noindent
where $\delta^8_{xy} \equiv \delta^4(x^\mu-y^\mu) \delta^4(\theta_x - \theta_y)$. From Eq.~(\ref{Effective_Propagator_V}) one can easily construct analytic expressions for the superdiagrams presented in Fig.~\ref{Figure:3:Three-points_2-loops_Polarization_C}.

Actually, in the limit of the vanishing external momenta contributions of all superdiagrams in Fig.~\ref{Figure:3:Three-points_2-loops_Polarization_C} to the function $ \mathcal{S} $ are equal to 0. This can be demonstrated using the same reasoning that was used in Ref.~\cite{Kuzmichev:2021yjo} for similar diagrams containing a loop of the superfileds $\phi_i$ or $\Phi_i$.

Namely, in the superdiagrams (c1), (c2), and (c3) the external gauge $\bar{D}^2 H$-line is attached to a quantum gauge superfield propagator. However, the resulting triple gauge vertex with an external $\bar{D}^2H $-line vanishes in the limit of its vanishing momentum, see Ref. \cite{Kuzmichev:2021yjo} for details.

\begin{figure}[h]
	\begin{picture}(0,3.4)
		\put(2.5,0){\includegraphics[scale=0.3,clip]{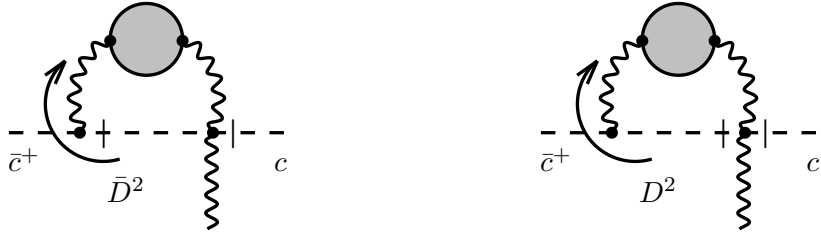}}
		\put(2.5,0.8){$\bar c^+$} \put(6,0.8){$c$} \put(3.7,1.2){$\bm{|}$} \put(3.8,0.4){$\bar D^2$} \put(5.4,1.2){$\bm{|}$}
		\put(9.5,0){\includegraphics[scale=0.3,clip]{3point_byparts.eps}}
		\put(9.5,0.8){$\bar c^+$} \put(13,0.8){$c$} \put(11.85,1.2){$\bm{|}$} \put(10.8,0.4){$D^2$} \put(12.4,1.2){$\bm{|}$}
	\end{picture}
	\caption{In the superdiagrams (c4), (c5), and (c6) we integrate by parts with respect to the derivatives $D^2$ and $\bar D^2$ coming from the ghost propagator. After this, in the limit of the vanishing external momenta they will act on the gauge superfield propagator and, in the end, on the transversal polarization operator.}
	\label{Figure:5:By_Parts}
\end{figure}

In the superdiagrams (c4), (c5), and (c6) one can integrate the covariant derivatives $D^2$ or $\bar D^2$ coming from the ghost propagators

\begin{equation}\label{Ghost_Propagator}
	\frac{D_x^2 \bar D_y^2}{4\partial^2} \delta^8_{xy} \qquad \mbox{or} \qquad \frac{\bar D_x^2 D_y^2}{4\partial^2} \delta^8_{xy}
\end{equation}

\noindent
by parts as illustrated in Fig.~\ref{Figure:5:By_Parts} (for the superdiagram (c6)) and take into account that in the limit of the vanishing external momenta no supersymmetric covariant derivatives can act on the external legs (except for $\bar{D}^2$ inside $\bar{D}^2 H$). (Otherwise, the result will be proportional to the vanishing external momenta.) Therefore, after integrating by parts the derivatives $D^2$ or $\bar D^2$ will act on a propagator of the quantum gauge superfield, and, eventually, on the polarization operator, which is transversal due to the Slavnov--Taylor identity (see Eq. (\ref{Dq_Definition})). Thus, the result vanishes because

\begin{equation}
	D^2 \partial^2\Pi_{1/2} = 0;\qquad \bar D^2 \partial^2\Pi_{1/2} = 0.
\end{equation}

\noindent
This implies that the superdiagrams (c4), (c5), and (c6) do not contribute to the function $ \mathcal{S} $ in the limit $p \to 0$, $q \to 0$.

\begin{figure}[h]
	\begin{picture}(0,3.2)		
		\put(3,0.3){\includegraphics[scale=0.37]{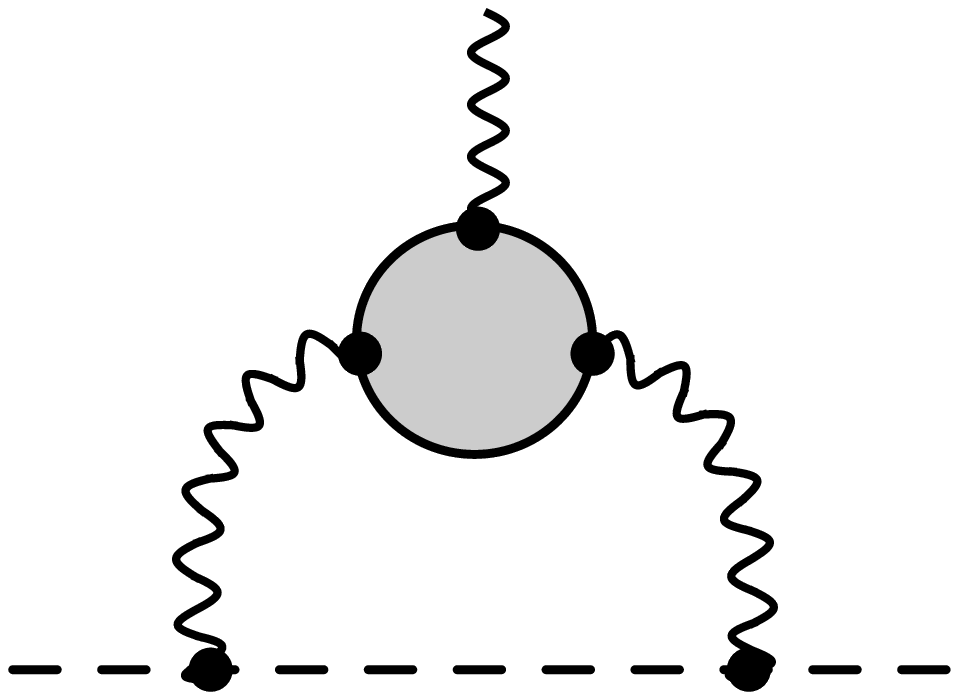}}	
		\put(2.8,2.7){(d1)}		
		\put(3,0){$ \bar{c}^+ $}
		\put(6.5,0){$ c$}
		\put(3.8,2.7){$ \bar{D}^2 H$}
		
		\put(10,0.3){\includegraphics[scale=0.37,clip]{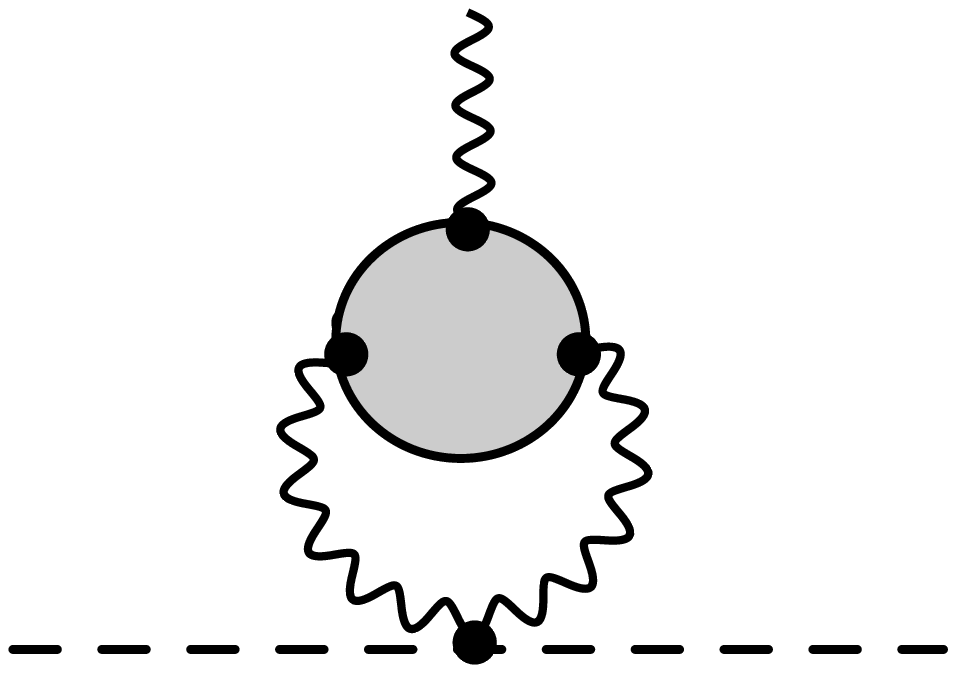}}		
		\put(10,2.7){(d2)}
	\end{picture}
\caption{Two-loop superdiagrams in which an external $ \bar{D}^2 H $-leg is attached to the one-loop polarization operator of the quantum gauge superfield denoted by a gray disk.}
\label{Figure:6:}
\end{figure}

\begin{figure}[h]
	\begin{picture}(0,8.4)
		\put(1,6){\includegraphics[scale=0.25,clip]{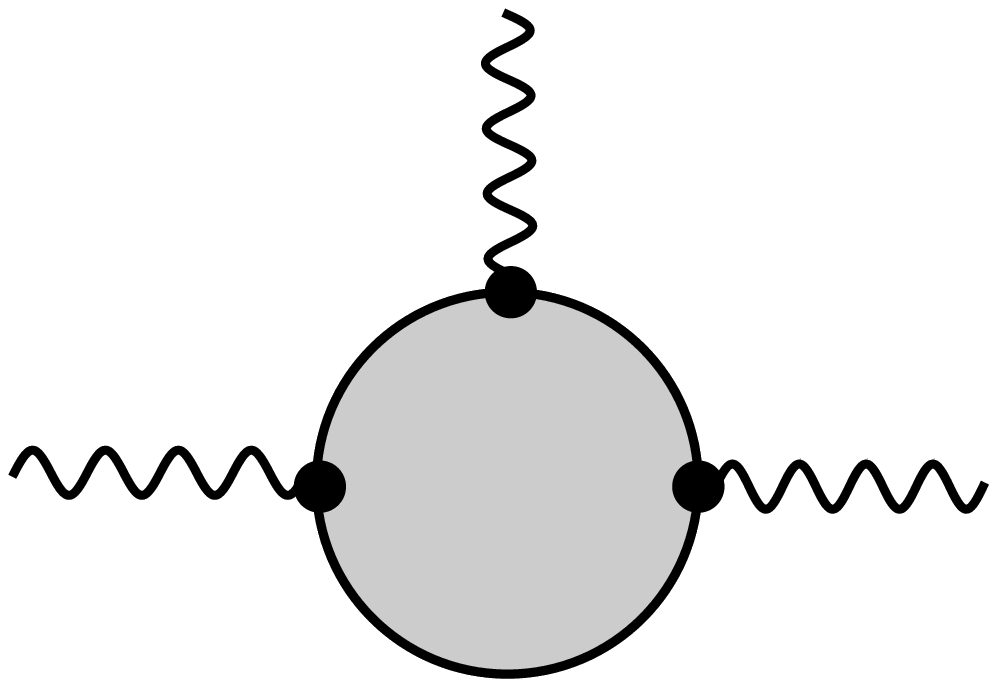}}
		\put(1.3,7.4){$ \bar{D}^2 H$}
		\put(4,6.4){$ \to $}
		\put(1,6.7){$ V $}
		\put(3.3,6.65){$ V $}

		\put(4.7,5.5){\includegraphics[scale=0.25,clip]{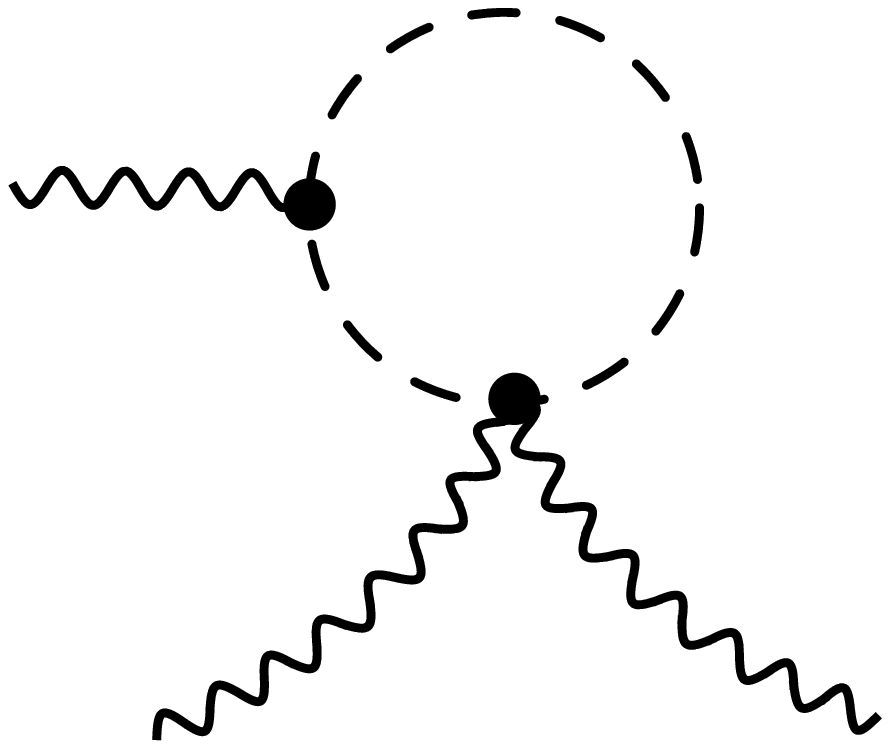}}
		\put(4.5,7.2){$ \bar{D}^2 H$}
		\put(5,5.8){$ V $}
		\put(6.7,5.8){$ V $}
		\put(7.2,6.4){$ + $}
		
		\put(8.2,5.9){\includegraphics[scale=0.25,clip]{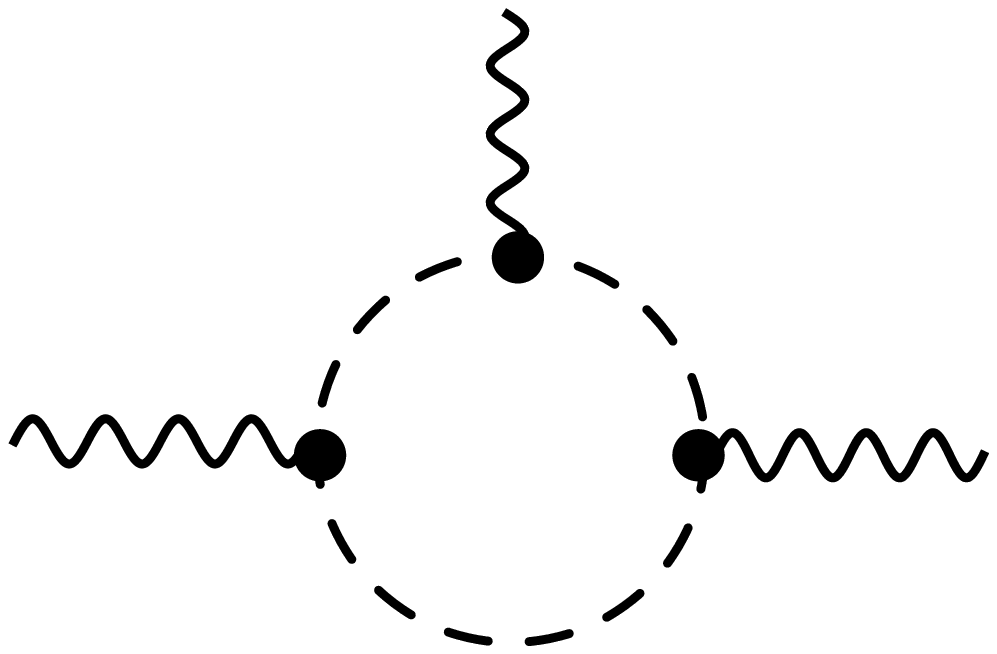}}
		\put(8.5,7.2){$ \bar{D}^2 H$}
		\put(8.2,6.6){$ V $}
		\put(10.5,6.55){$ V $}
		\put(11.4,6.4){$ + $}
		
		\put(12.4,5.9){\includegraphics[scale=0.25,clip]{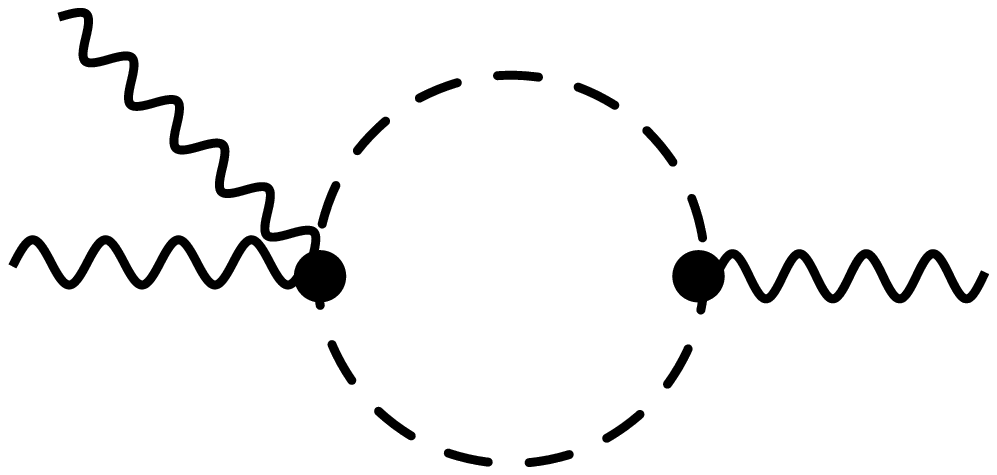}}
		\put(12.4,7.2){$ \bar{D}^2 H$}
		\put(12.2,6.55){$ V $}
		\put(14.6,6.55){$ V $}
		\put(1,3.7){$ + $}
		\put(1.4,2.9){\includegraphics[scale=0.25,clip]{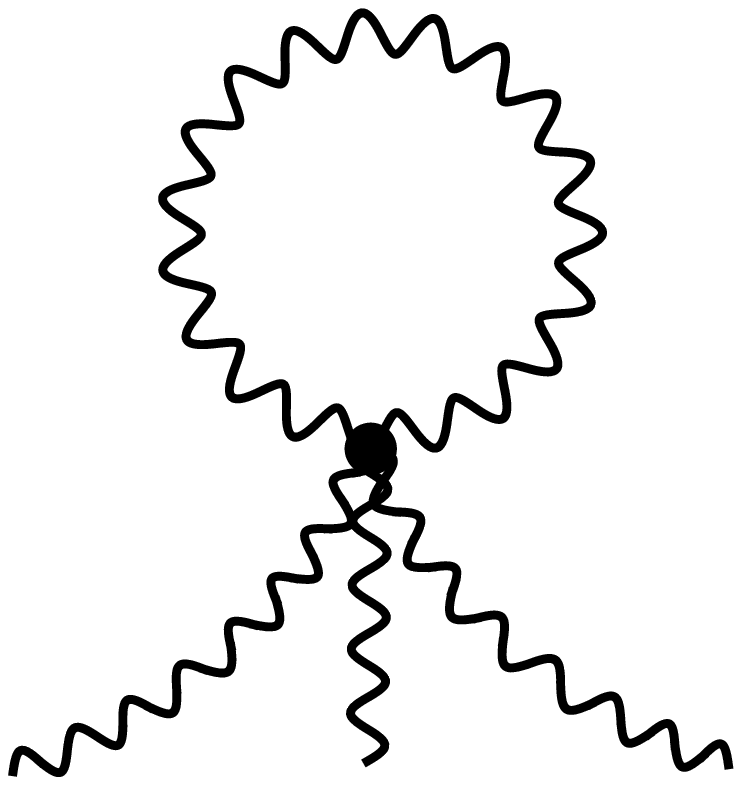}}
		\put(4,3.7){$ + $}
		
		\put(4.7,2.9){\includegraphics[scale=0.25,clip]{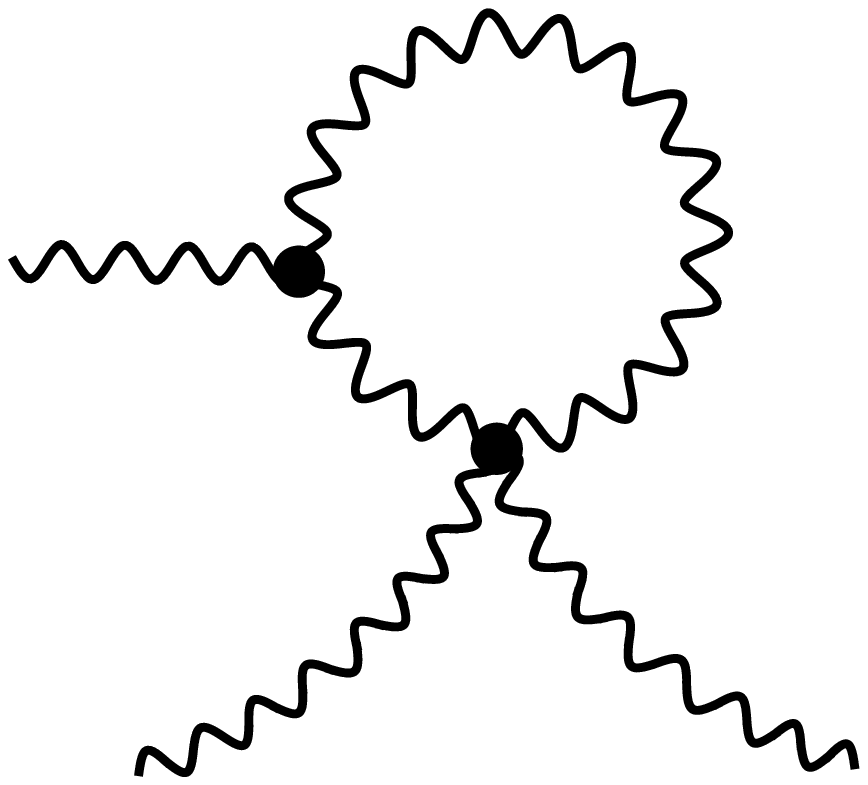}}
		\put(7.2,3.7){$ + $}
		
		\put(8.2,3.2){\includegraphics[scale=0.25,clip]{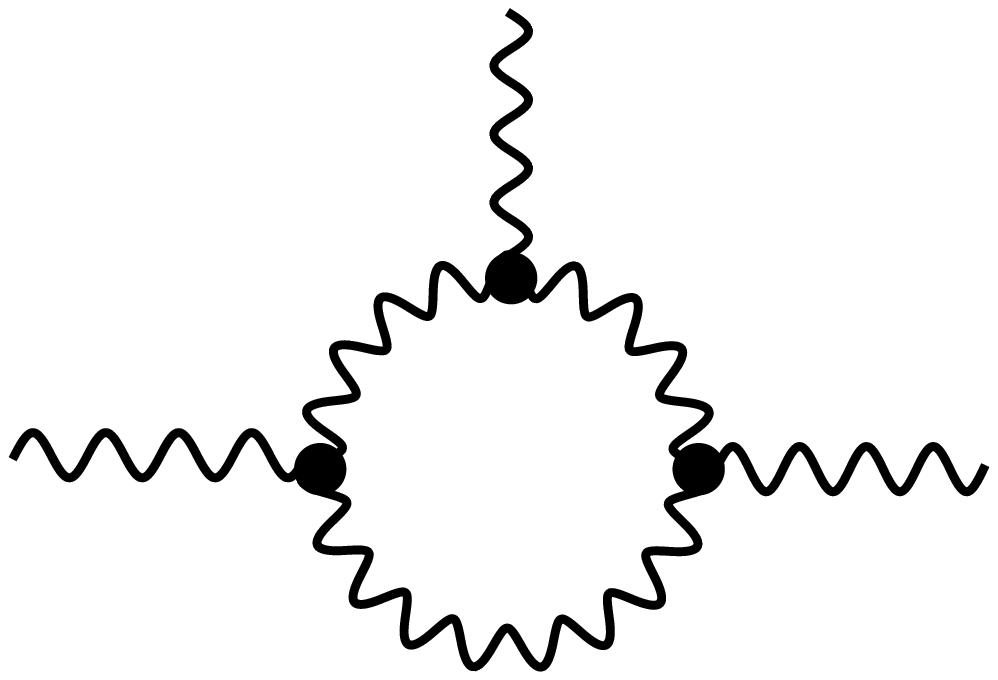}}
		\put(11.4,3.7){$ + $}
		
		\put(12.4,3.2){\includegraphics[scale=0.25,clip]{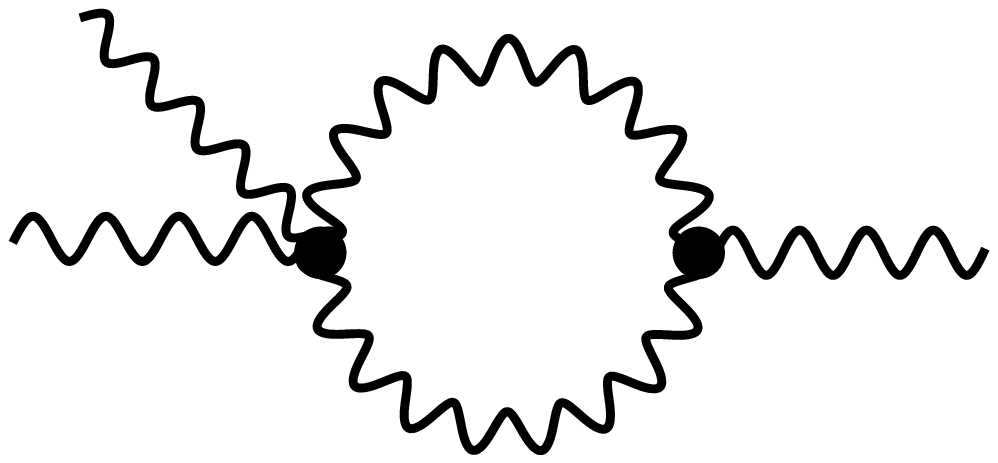}}
	
		\put(1,1.1){$ + $}
		\put(1.4,0.3){\includegraphics[scale=0.25,clip]{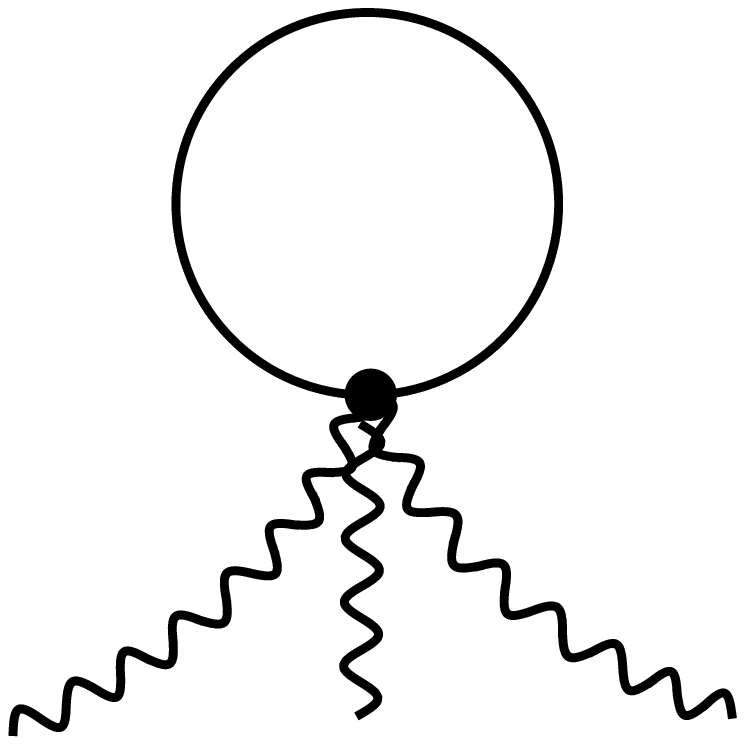}}
		\put(4,1.1){$ + $}
		
		\put(4.7,0.3){\includegraphics[scale=0.25,clip]{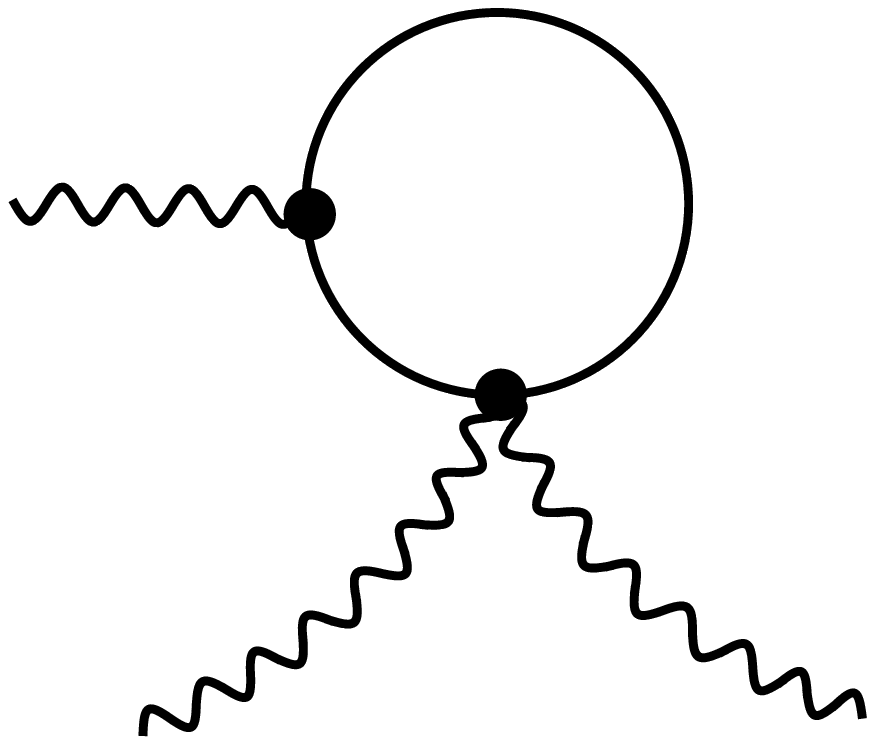}}
		\put(7.2,1.1){$ + $}
		
		\put(8.2,0.6){\includegraphics[scale=0.25,clip]{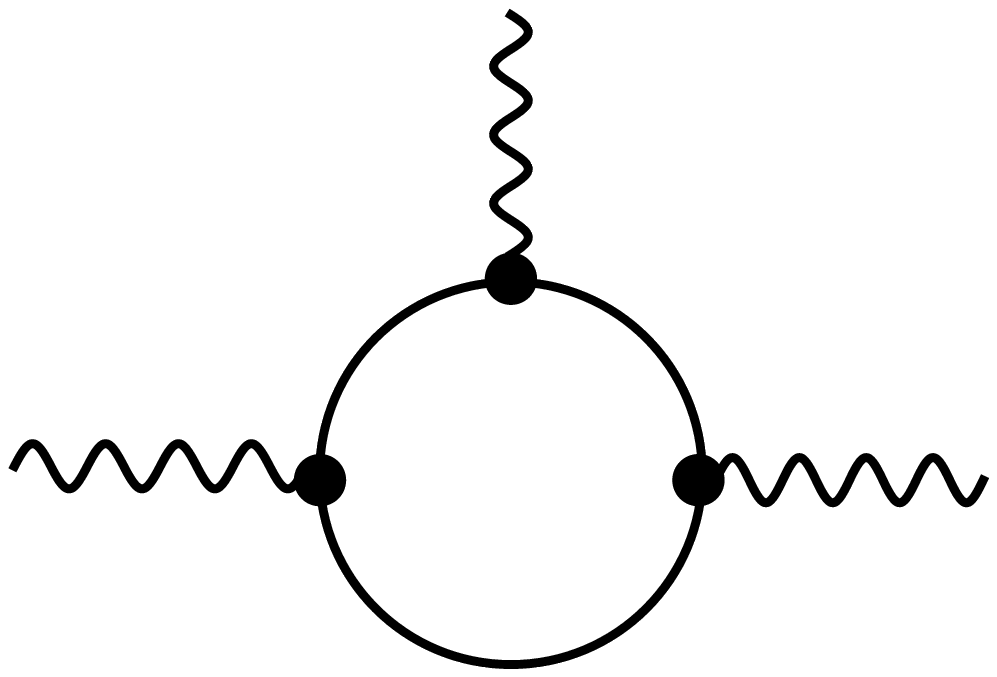}}
		\put(11.4,1.1){$ + $}
		
		\put(12.4,0.6){\includegraphics[scale=0.25,clip]{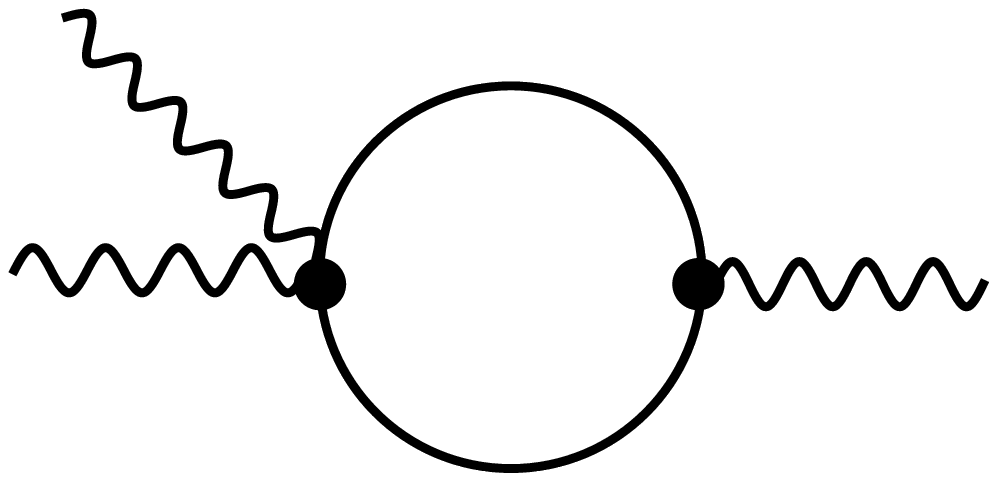}}
	\end{picture}
	\caption{Subdiagrams obtained by attaching an external $ \bar{D}^2 H $-leg to a part of the one-loop polarization operator of the quantum gauge superfield containing a loop of the Faddeev--Popov ghosts, a loop of the quantum gauge superfield, or a loop of the Pauli--Villars superfields $ \varphi_{1,2,3} $.}
	\label{Figure:7:}
\end{figure}

4. The superdiagrams of the fourth group are presented in Fig.~\ref{Figure:6:}. They contain all possible attachments of an external $\bar{D}^2 H$-leg to a part of the one-loop polarization operator in which the loop  corresponds to the quantum gauge superfield, to the Faddeev--Popov ghosts, or to the Pauli--Villars superfields $\varphi_{1,2,3}$. (A part containing a loop of the matter superfields or a loop of the Pauli--Villars superfields $\Phi_i$ has already been considered in Ref. \cite{Kuzmichev:2021yjo}.) The subdiagrams obtained after the above described attachments of an external $\bar{D}^2 H$-leg are depicted in Fig.~\ref{Figure:7:}.
After a rather nontrivial calculation we have obtained that in the limit $p\to 0$ (where $-p^\mu$ is the momentum of the superfield $\bar D^2 H$) their sum is equal to 0, although some individual supergraphs do not vanish. This implies that the fourth group of the superdiagrams (presented in Fig.~\ref{Figure:6:}) does not contribute to the function ${\cal S}$ in the limit of the vanishing external momenta.

\begin{figure}[h]
	\begin{picture}(0,3)		
		\put(1.1,0){\includegraphics[scale=0.12,clip]{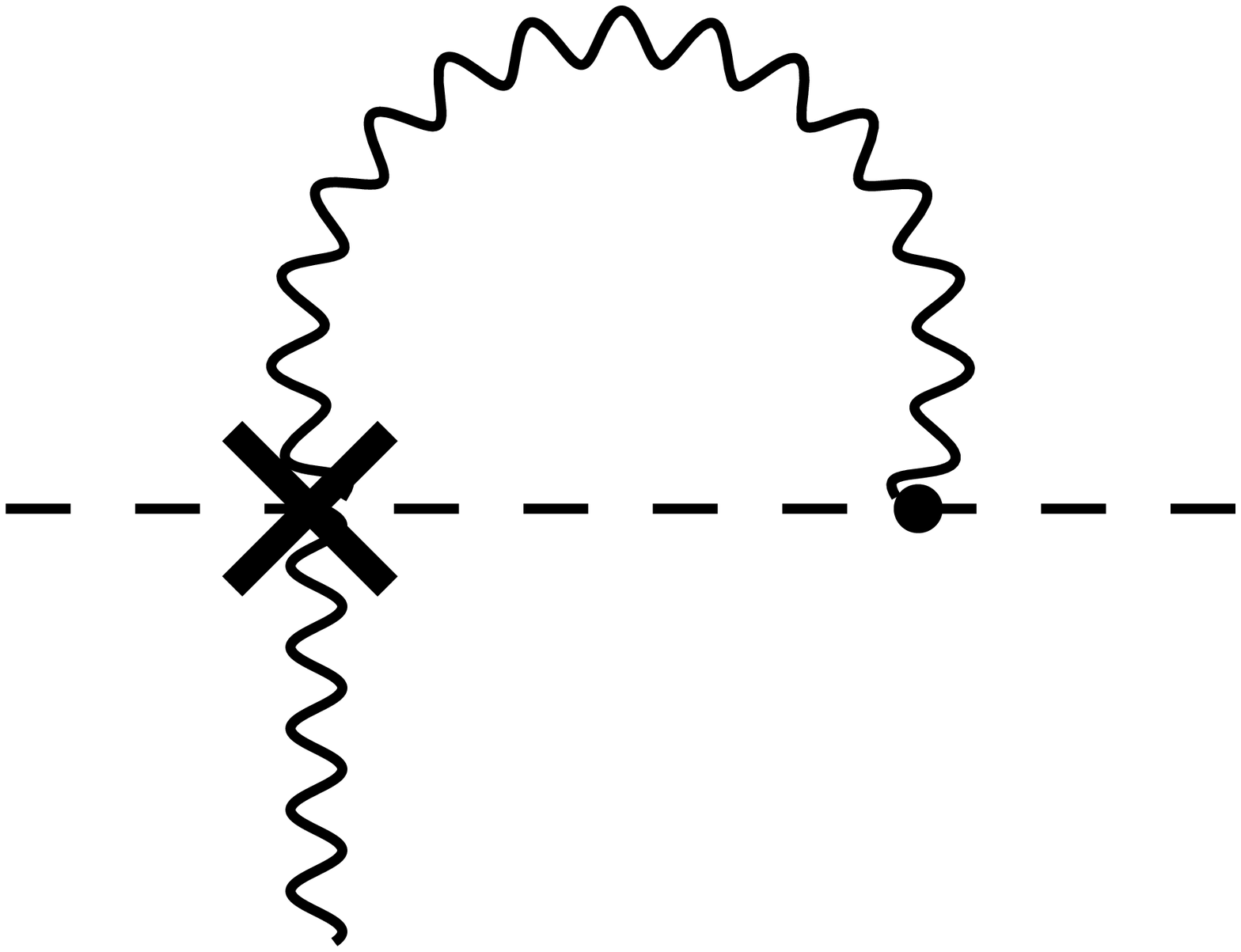}}	
		\put(1,2){(e1)}	
		
		\put(0.9,0.4){$ \bar{c}^+ $}
		\put(3.1,0.4){$ c$}
		\put(1.3,-0.5){$ \bar{D}^2 H$}
		
		\put(5.1,0){\includegraphics[scale=0.12,clip]{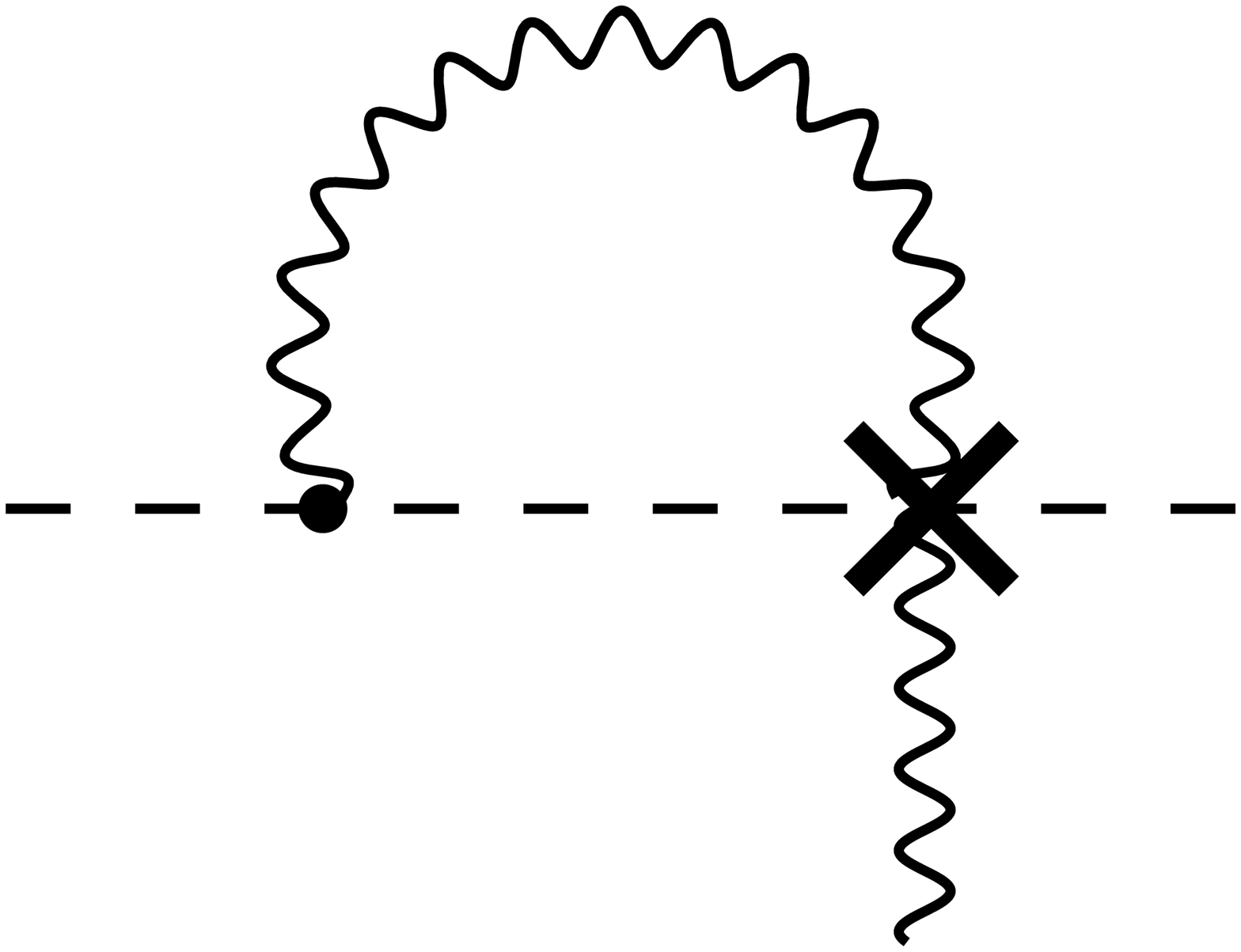}}		
		\put(5,2){(e2)}
			
		\put(8.8,-0.11){\includegraphics[scale=0.13,clip]{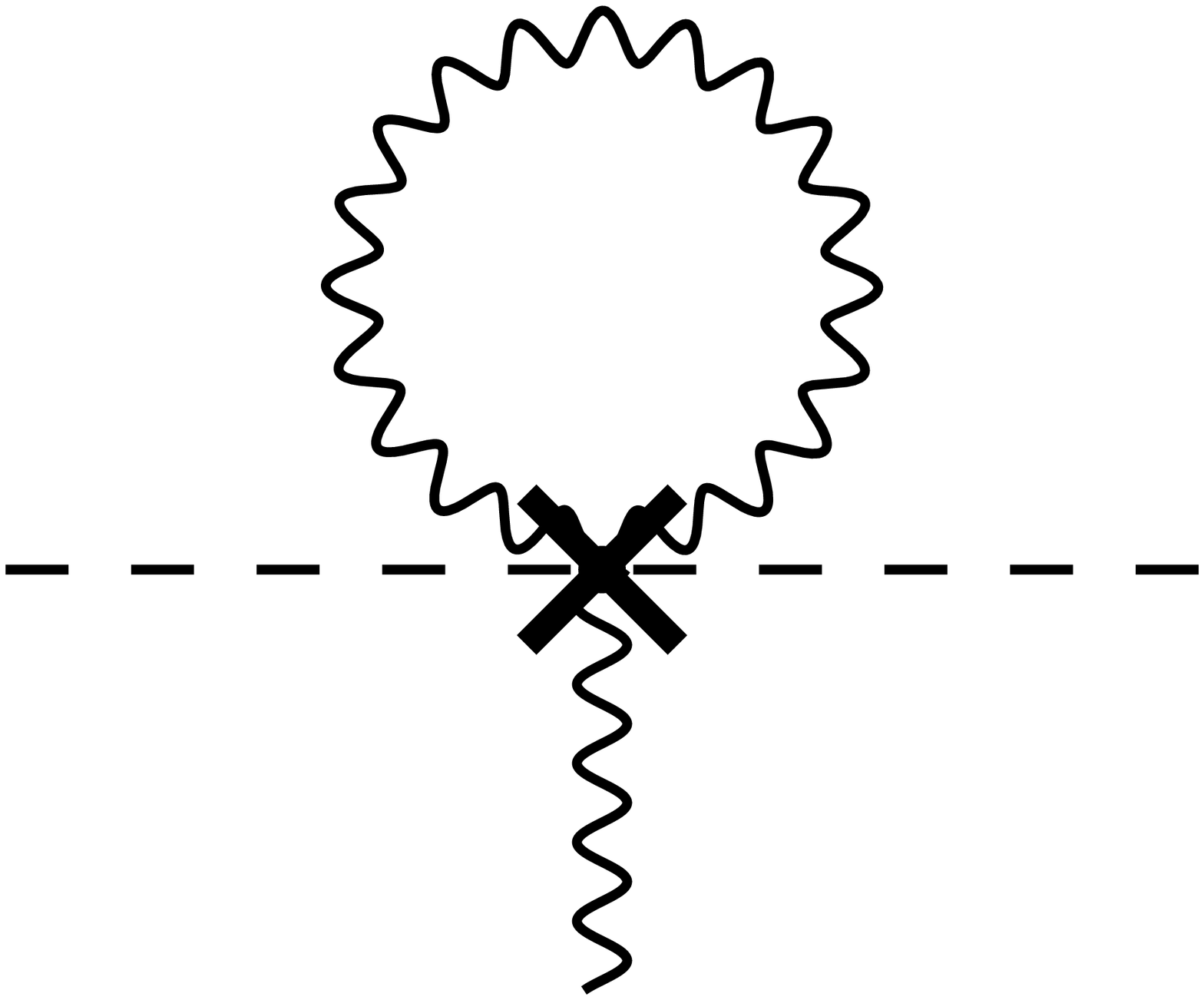}}	
		\put(8.8,2){(e3)}
		
		\put(12.7,0.56){\includegraphics[scale=0.13,clip]{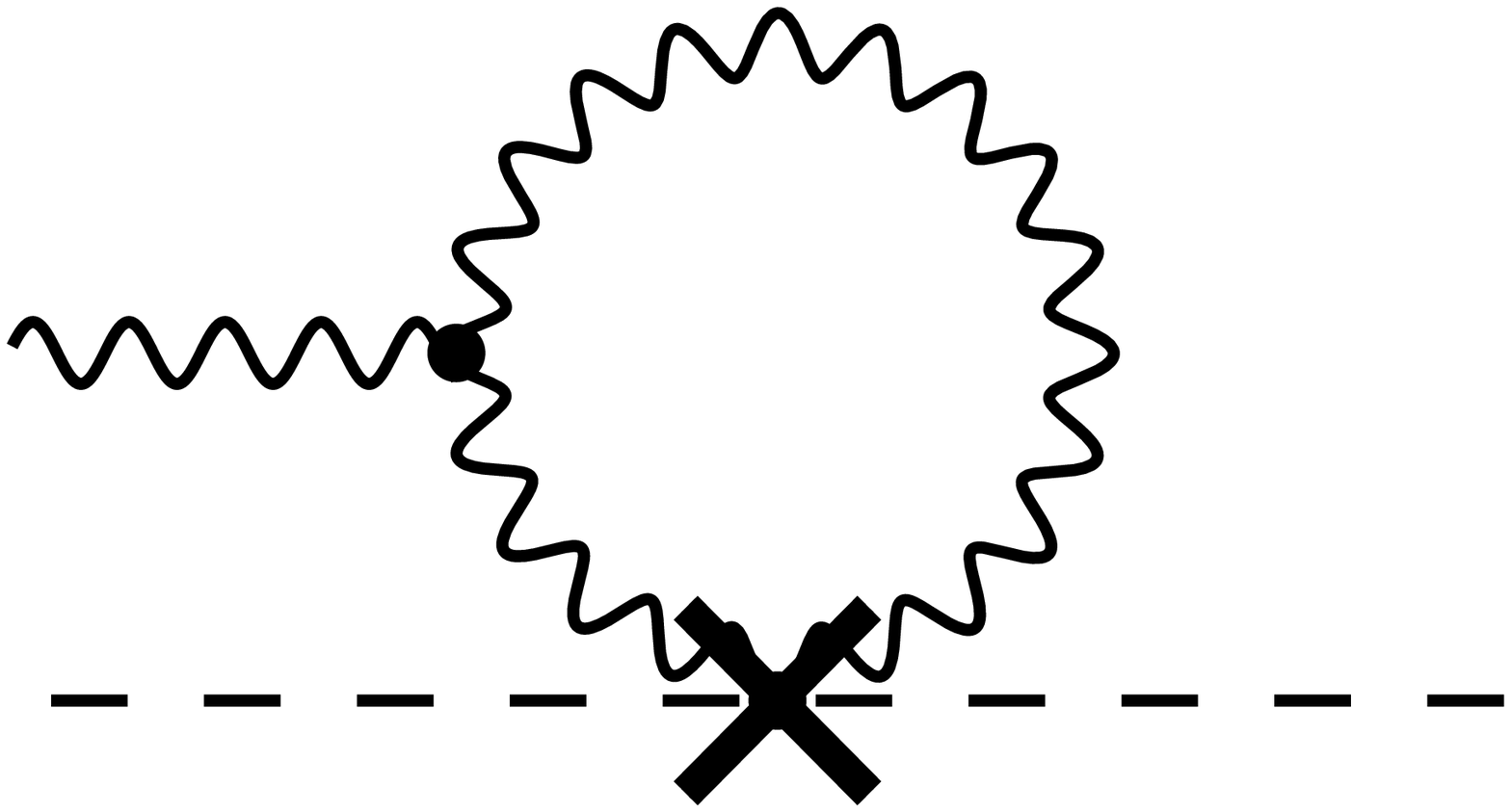}}
		\put(12.7,2){(e4)}		
	\end{picture}
	\vspace*{5mm}
	\caption{One-loop superdiagrams containing vertices which come from the nonlinear terms in the function $\mathcal{F}(V)$. These vertices are marked by crosses.}
	\label{Figure:8:}
\end{figure}

5. The last fifth group of the superdiagrams is presented in Fig.~\ref{Figure:8:}. They include vertices containing parameters describing the nonlinear renormalization of the quantum gauge superfield. Taking into account that the function $\mathcal{F}(V)$ can contain only odd powers of $V$ \cite{Piguet:1984mv} we see that the only term in the function $\mathcal{F}(V)^A$ relevant in the considered approximation is $e_0^2\, y_0\, G^{ABCD} V^B V^C V^D$. From Eq. (\ref{Action_Faddeev-Popov_Ghosts}) we see that the (tree) ghost-gauge vertices generated by this term can be written as

\begin{eqnarray}
    &&\hspace*{-5mm} \int d^4x\, d^4\theta\, \Big\{ -\frac{3}{4} e_0^2\, y_0\, G^{ABCD} (\bar c^A + \bar c^{+A}) V^C V^D (c^B-c^{+B}) - \frac{i}{4}e_0^3 \, y_0 \, \left[ f^{ABK}G^{KCDE} \right.
    \nonumber\\
    &&\hspace*{-5mm} \left. +3 f^{EBK}G^{KACD} \right] (\bar c^A + \bar c^{+A}) V^C V^D V^E (c^B+c^{+B})\Big\}.
\end{eqnarray}

\noindent
Making the replacement $V \to V + \bar{D}^2 H$ and extracting terms linear in $\bar D^2 H$ we see that the vertices with crosses in the considered superdiagrams are given by the expression

\begin{eqnarray}
	&&\hspace*{-5mm} \int d^4x\, d^4\theta\, \Big\{ -\frac{3}{2} e_0^2\, y_0\, G^{ABCD} (\bar c^A + \bar c^{+A}) \bar{D}^2 H^C V^D (c^B-c^{+B}) - \frac{3i}{4}e_0^3 \, y_0 \, \left[ f^{ABK}G^{KCDE}   \right.
	\nonumber\\
	&&\hspace*{-5mm} \left. + 2 f^{EBK}G^{KACD} + f^{CBK}G^{KAED} \right] (\bar c^A + \bar c^{+A}) \bar{D}^2 H^C V^D V^E (c^B+c^{+B})\Big\}.
\end{eqnarray}

Now we can analyze the contributions of the superdiagrams (e1) --- (e4). The superdiagrams (e1) and (e2) include the factor

\begin{equation}
	G^{ABDE} f^{CDE} = 0.
\end{equation}

\noindent
In this expression the indices $A$, $B$, and $C$ correspond to the external lines. The result vanishes because the totally symmetric tensor $G^{ABDE}$ is multiplied by the totally antisymmetric structure constants $f^{CDE}$. Similarly, the superdiagram (e3) vanishes because it is proportional to

\begin{equation}
    f^{ABK}G^{KCDD} + 2 f^{DBK}G^{KACD} + f^{CBK} G^{KADD} = \frac{5}{6} (C_2)^2 \left( f^{ABC} + f^{CBA} \right) =0,	
\end{equation}

\noindent
where we took into account that $G^{KCDD} = 5 (C_2)^2 \delta^{KC}/6$. The last superdiagram (e4) contains the triple gauge vertex with an external gauge $ \bar{D}^2 H $-line considered in the limit of the vanishing $\bar D^2 H$ momentum. As we have already mentioned, all superdiagrams containing this vertex are equal to 0 in this limit, see Ref. \cite{Kuzmichev:2021yjo} for details. Therefore, all superdiagrams depicted in Fig. \ref{Figure:8:} vanish in the limit $p\to 0$, $q\to 0$.

Thus, we see that the total contribution of all groups of superdiagrams is equal to 0 in the limit of the vanishing external momenta. Due to the renormalizability of the considered theory this implies that the two-loop quantum correction to the function $ \mathcal{S}(p,q)$ proportional to $(C_2)^2$ is UV finite for nonvanishing external momenta, although some individual superdiagrams are UV divergent. Taking into account that the $C_2 T(R)$ contribution is also finite in the UV region, we conclude that the triple gauge-ghost vertices are not renormalized in the two-loop approximation in exact agreement with the general statement proved in \cite{Stepanyantz:2016gtk}.

\section{Conclusion}
\hspace*{\parindent}

By an explicit calculation we have verified that the sum of the two-loop quantum corrections to the triple gauge-ghost vertices in renormalizable $ \mathcal{N} =1 $ supersymmetric gauge theories regularized by higher covariant derivatives is finite in the ultraviolet region. Actually, in this paper we have demonstrated the finiteness of the contribution proportional to $(C_2)^2$, while the finiteness of the $C_2 T(R)$ contribution (which comes from the two-loop superdiagrams containing a matter loop) has been checked in Ref.~\cite{Kuzmichev:2021yjo}. The calculation was done with the help of the ${\cal N}=1$ superspace formalism for the general $\xi$ gauge fixing condition. (Some parameters similar to the gauge parameter $\xi$ also appear due to the necessity to take into account the nonlinear renormalization of the quantum gauge superfield.) The results obtained in this paper and  in Ref.~\cite{Kuzmichev:2021yjo} confirm the nonrenormalization theorem for the triple gauge-ghost vertices which was derived in all orders of the perturbation theory in Ref.~\cite{Stepanyantz:2016gtk}, see also Ref.~\cite{Korneev:2021zdz}, where the proof has been extended to the case of theories with multiple gauge couplings. Strange as it may seem, the ultraviolet finiteness of the triple gauge-ghost vertices plays an important role for understanding the perturbative origin of the exact NSVZ $\beta$-function and is a key ingredient needed for its derivation, see Refs. \cite{Stepanyantz:2016gtk,Stepanyantz:2019ihw,Stepanyantz:2019lfm,Stepanyantz:2020uke}. Therefore, the two-loop perturbative check made in this paper also confirms some general statements needed to understand the behaviour of supersymmetric theories at the quantum level.

\section*{Acknowledgments}
\hspace*{\parindent}

This work was supported by Foundation for Advancement of Theoretical Physics and Mathematics ``BASIS'', grants  No. 19-1-1-45-5 (M.K.), 18-2-6-159-1 (N.M.), 18-2-6-158-1 (S.N.), 19-1-1-45-3 (I.S.), and 19-1-1-45-1 (K.S.).

\end{document}